\documentclass[aps,twocolumn,superscriptaddress,floatfix,prl,10pt]{revtex4-2}

\usepackage{graphicx}
\usepackage[utf8]{inputenc}
\usepackage{amssymb}
\usepackage{amsmath}
\usepackage[colorlinks,allcolors=blue]{hyperref}
\usepackage{braket}
\usepackage{placeins}

\begin{document}

    \title{Disorder-Enhanced and Disorder-Independent Transport with Long-Range Hopping: Application to Molecular Chains in Optical Cavities}

\author{Nahum C. Ch\'avez}
\affiliation{Dipartimento di Matematica e Fisica and Interdisciplinary Laboratories for Advanced Materials Physics, Universit\`a Cattolica, via Musei 41, 25121 Brescia, Italy}
\affiliation{Benem\'erita Universidad Aut\'onoma de Puebla, Apartado Postal J-48, Instituto de F\'isica,  72570, Mexico}

\author{Francesco Mattiotti}
\affiliation{Dipartimento di Matematica e Fisica and Interdisciplinary Laboratories for Advanced Materials Physics, Universit\`a Cattolica, via Musei 41, 25121 Brescia, Italy}
\affiliation{Istituto Nazionale di Fisica Nucleare,  Sezione di Pavia, via Bassi 6, I-27100, Pavia, Italy}
\affiliation{Department of Physics, University of Notre Dame, Notre Dame, Indiana 46556, USA}

\author{J. A. M\'endez-Berm\'udez}
\affiliation{Benem\'erita Universidad Aut\'onoma de Puebla, Apartado Postal J-48, Instituto de F\'isica,  72570, Mexico}

\author{Fausto Borgonovi}
\affiliation{Dipartimento di Matematica e Fisica and Interdisciplinary Laboratories for Advanced Materials Physics, Universit\`a Cattolica, via Musei 41, 25121 Brescia, Italy}
\affiliation{Istituto Nazionale di Fisica Nucleare,  Sezione di Milano, via Celoria 16, I-20133, Milano, Italy}

\author{G. Luca Celardo}
\affiliation{Benem\'erita Universidad Aut\'onoma de Puebla, Apartado Postal J-48, Instituto de F\'isica,  72570, Mexico}

\begin{abstract}
Overcoming   the detrimental effect of disorder at the nanoscale  is very hard since disorder induces localization and an exponential suppression of transport efficiency. Here we unveil novel and  robust quantum transport regimes achievable in nanosystems by exploiting long-range hopping.
We demonstrate that in a 1D disordered nanostructure in the presence of long-range hopping,  transport efficiency, after decreasing exponentially with disorder at first, is then  enhanced by disorder [disorder-enhanced transport (DET) regime] until, counterintuitively, it reaches a disorder-independent transport (DIT) regime, persisting over several orders of disorder magnitude in realistic systems.  
To enlighten the  relevance of our results, we demonstrate that  an ensemble of emitters in a cavity  can be described by an
effective long-range Hamiltonian.
The specific case of a disordered molecular wire placed in an optical cavity is discussed, showing that the DIT and DET regimes can be reached with state-of-the-art experimental setups.
\end{abstract}


\maketitle

{\it Introduction.---}Achieving high efficiency for energy or charge transport in quantum wires is  fundamental for quantum technologies related to quantum computation and basic energy science~\cite{coles2014,vidal,pupillo-old,dyes,highsp,mineo,du,rubio,expcavity,pupillo-charge,pupillo-fermions}.  One of the main challenges  is to control the detrimental effects of  noise and disorder which naturally occur in realistic situations. It is well known that disorder induces localization~\cite{Anderson,Abrahams10} and exponential suppression of transport in typical 1D 
nanostructures. One of the most ambitious goals in quantum transport is to achieve dissipationless quantum wires, able to transport energy or charge without suffering the detrimental effects of disorder and/or noise.
  
\begin{figure}[ht!]
    \centering
    \includegraphics[width=\columnwidth]{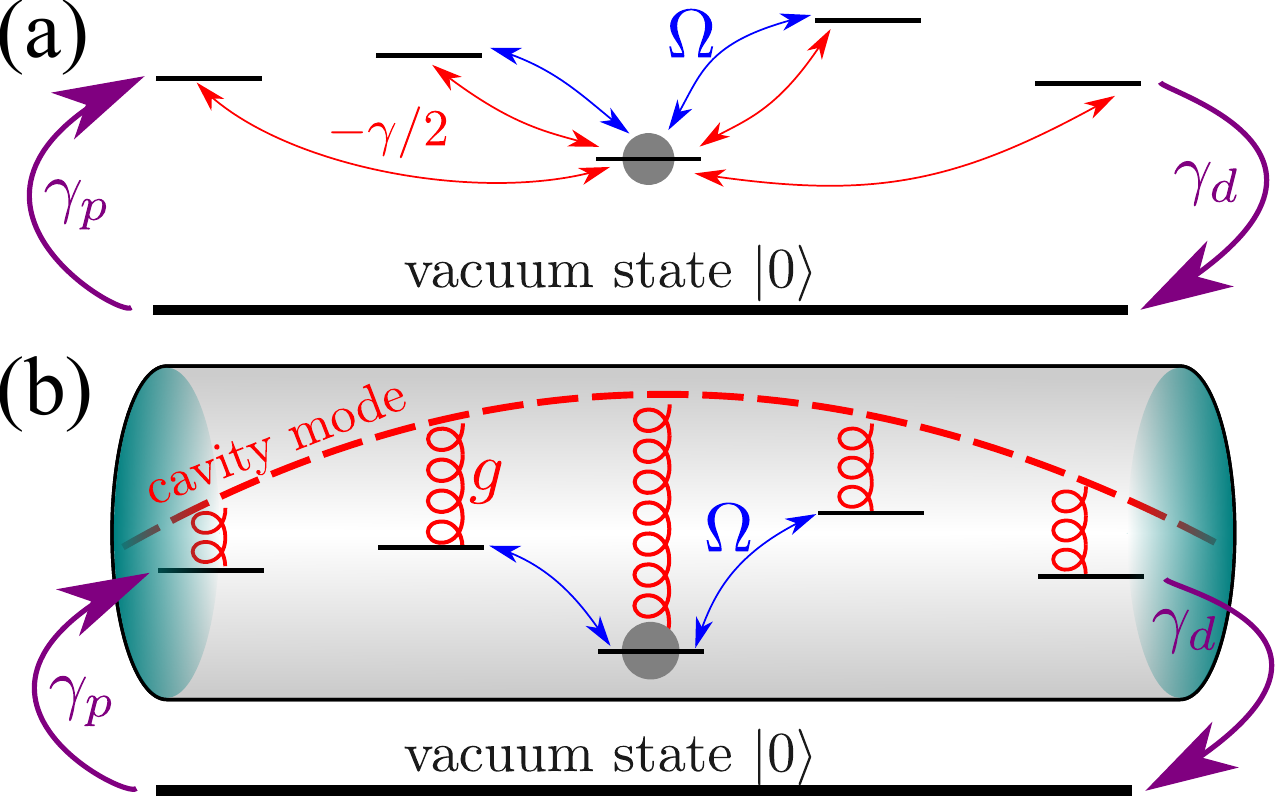}
    \caption{(a,b) Two different setups for a disordered chain with excitation pumping $\gamma_p$ at one edge of the chain and draining $\gamma_d$ at the opposite edge. Here, $\Omega$ is the hopping between nearest-neighbor sites. The arrows indicate the hopping paths available for an excitation (gray circle) present at the center of the chain. The energy of the sites is disordered. 
    (a) A long-range coupling $-\gamma/2$ is present between each pair of sites. (b) The chain is placed inside an optical cavity, where $g$ is the coupling of each site to the cavity mode.
    }
    \label{fig:0}
\end{figure}
Here, to overcome disorder suppression of transport,  we propose  to  exploit long-range interactions. 
Long-range interactions  can arise due to microscopic interactions or by engineering the coupling to  external degrees of freedom. They have been recently emulated  in ion traps~\cite{iontraps} and are relevant in several realistic systems such as cold atomic clouds~\cite{robin2017} and excitonic transport in molecular aggregates~\cite{schulten,Gulli2019,mukamelspano}. 
Long-range interactions present many contradictory features~\cite{robincs,lea2016,Kastner2015}. Specifically, the interplay of localization and long-range interactions is widely debated in literature~\cite{robincs,lea2016,duality2018,correlation2019,nosov2019,adame2005,Gorshkov2019,Silva2019,Mori2019}.
Indeed,  contrary  to  the  common  lore  that  long range should  destroy  Anderson  localization~\cite{levitov1989,mirlin}, strong signatures of localization  have been reported recently in long-range interacting systems~\cite{robincs,duality2018,correlation2019}, thus questioning their utility in achieving efficient transport.
Here we demonstrate that localized states in long-range interacting systems have a hybrid character, with an exponentially localized peak and extended tail, which allows these states to support robust quantum transport. 

Among the most important features of long-range systems, there is the emergence of a gapped ground state~\cite{nahum,robincs}. In the gapped regime, while the ground state is extended and robust to disorder, the excited states present  a hybrid nature  with an exponentially localized peak superimposed to an extended tail~\cite{alberto,alberto2013,robincs}.
  While being  very relevant to transport, since they constitute  the vast majority of the states, 
due to their hybrid nature  it is not clear what kind of transport they will be able to support. 
By using different standard figures of merit  of transport efficiency,  we unveil several regimes directly determined by the hybrid nature of the excited states. 
Specifically we develop a new method to compute the stationary current, based on an effective non-Hermitian Hamiltonian formalism, which is equivalent to a Lindblad master equation and it is much more efficient.
We demonstrate, in presence of  long-range hopping, the emergence of extremely robust transport regimes arising as the disorder strength is increased: a disorder-enhanced transport (DET) regime and, at larger disorder strength, a disorder-independent transport (DIT) regime, where transport efficiency is independent of disorder over several orders of magnitude of disorder strength. The latter regime persists until disorder is so large to close the energy gap. 
We can explain the origin of this interesting behavior by considering that in the presence of an energy gap, disorder will mix the excited states, while leaving the ground state  fully extended. The presence of an extended ground state 
imposes an orthogonality condition on the excited states which prevents their full single-site localization and generates an extended tail able to  support robust transport over the whole energy spectrum. 

In order to highlight the relevance of our findings, we analyze realistic setups consisting of an ensemble of emitters inside a cavity, focusing on the case of molecular chains in optical cavities. Recently these systems have been studied experimentally~\cite{expcavity} and analyzed theoretically~\cite{pupillo-old,vidal,pupillo-new}. Here we show that, in the strong coupling regime~\cite{SCcavity1,ultrastrong2013},  the cavity  induces  an effective long-range hopping between the emitters, allowing us to test our findings of both DET and DIT regimes in state-of-the-art experimental setups.

{\it Model.---}As a paradigmatic model of a disordered chain in the presence of long-range hopping,  
 we analyze the 1D Anderson model~\cite{Anderson} with  all-to-all hopping~\cite{robincs}, see Fig.~\ref{fig:0}(a),
\begin{equation}
H=H_0+V \quad \mbox{with} \quad V= -\frac{\gamma}{2} \sum_{i \ne j} | i \rangle \langle j|,
\label{HAcp}
\end{equation}
where $|j\rangle$ is the  site basis and  $\gamma$ is the strength of the distance-independent long-range hopping. $H_0$ describes the Anderson model where  
a particle  hops between neighbor sites of a linear chain in the presence of on-site disorder, %
\begin{equation}
H_0 = \sum_{j=1}^N \epsilon_j \ket{j}\bra{j} + \Omega \sum_{j=1}^{N-1} ( \ket{j}\bra{j+1}+\ket{j+1}\bra{j}) \, ,
\label{Ho}
\end{equation}
where $\epsilon_i$ are random energies uniformly distributed in $[-W/2, W/2]$, where $W$ is the disorder strength and $\Omega$ is the tunneling transition amplitude between nearest neighbor sites. 

 The eigenstates of the Anderson Model ($\gamma=0$) are localized exponentially, $\psi_n \sim \exp(-|n-n_0|/\xi)$, where  $\xi \approx 105.2(\Omega/W)^2$ is
 the localization length in the middle of the energy band. 
This implies  that the transmission always decays exponentially with the disorder strength as $\approx \exp(-N/\xi)$~\cite{Thouless79,Izrailev98}.

%
\begin{figure*}
  \centering
\includegraphics[width=\textwidth]{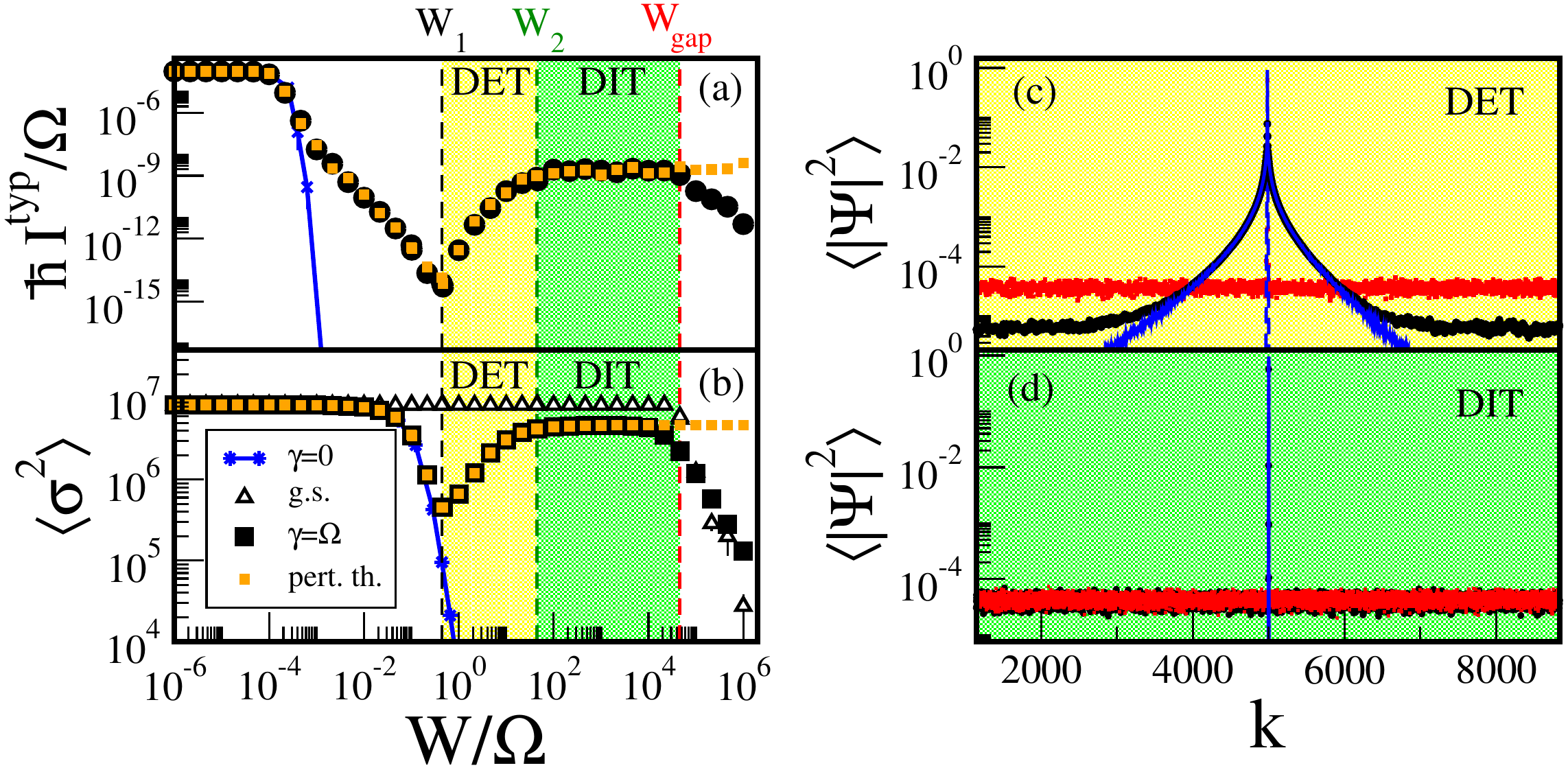}
  \caption{(a) Normalized typical  current $\hbar I^{\rm typ}/\Omega$ versus the normalized static disorder $W/\Omega$. 
  (b) Average variance  $\braket{\sigma^2}$ versus the normalized static disorder $W/\Omega$ for the ground state (triangles) and the excited states (all other sets).   The blue curves show the case $\gamma=0$, the orange squares the perturbative approach (details are given in the SM). Dashed vertical lines indicate the different critical disorders given by Eqs.~\eqref{WGAP},~\eqref{W1} and~\eqref{W2}. (c),(d) Average shape of the eigenfunctions (details concerning the process of averaging are given  in SM), $\braket{|\Psi|^2}$ versus the site basis $k$. Different disorder regimes are shown.
   (c) (DET)  $W_1 \le W  \le W_2$, $W/\Omega=1$ (black) and $W/\Omega=44.02$ (red);
   (d) (DIT)  $W_2 < W < W_{\rm gap}$, $W/\Omega=10^2$ (black) and  $W/\Omega=10^3$ (red). 
   Here, $N=10^4$, $\gamma_p=\gamma_d=\gamma=\Omega$, and $N_r=100$ disorder configurations.
   In (c),(d) symbols are compared with  blue curves indicating the case $\gamma=0$. }
  \label{Fig:1}
\end{figure*}

In the presence of long-range hopping ($\gamma \ne 0$), and in the absence of disorder ($W=0$), the emergence of an energy gap $\Delta=N\gamma/2$ has been found in Refs.~\cite{nahum,robincs}.
Indeed, the long-range hopping induces the fully symmetric ground state to be gapped from the other excited states. Disorder will  destroy the energy gap above the disorder threshold~\cite{nahum} [for details see the Supplementary Material (SM)~\footnote{See Supplemental Material for detailed calculations and supplemental results, including 
Refs.~\cite{Luca09,datta,sokolov}.}],
%
\begin{equation}\label{WGAP}
    W_{\rm gap}=\frac{\gamma}{2} N \ln{N}.
\end{equation}
In order to understand how transport properties are affected by long-range hopping,
we analyze several  figures of merit of transport efficiency, focusing on  the stationary current widely used in literature~\cite{pupillo-new,pupillo-old,vidal}. Pumping and draining are introduced at the chain edges, see Fig.~\ref{fig:0}(a) and the dynamics is described by the Lindblad master equation~\cite{petruccione}
\begin{equation}
    \frac{d\rho}{dt} = -\frac{i}{\hbar} [ H, \rho ] + \sum_{\eta=p,d} {\cal L}_\eta[\rho] \, ,
    \label{master}
\end{equation}
where  ${\cal L}_\eta[\rho] = -\{ L^\dag_\eta L_\eta, \rho \} + 2L_\eta \rho L^\dag_\eta$ are two dissipators inducing pumping on the first site \mbox{[$L_p=\sqrt{\gamma_p/(2\hbar)}\ket{1}\bra{0}$]} and draining from the last site [$L_d=\sqrt{\gamma_d/(2\hbar)}\ket{0}\bra{N}$], respectively ($\ket{0}$ is the vacuum state).
 From the steady-state solution of Eq.~\eqref{master} one can  find the stationary  current,
\begin{equation}
    I=\frac{\gamma_d}{\hbar} \bra{N}\rho_{SS}\ket{N} \, ,
    \label{Io}
\end{equation}
where $\rho_{\rm SS}$ is the steady-state density operator. 
Since the master equation approach is numerically very expensive, we use a definition of current based on  a non-Hermitian Schr\"odinger equation,  computationally less expensive. The results obtained with this approach are identical to 
the master equation method, as we prove analytically in Sec.~V.B of the SM.
To define the current, we compute the average time needed to leave the 1D chain if the excitation is initially on the first site $\ket{1}$ and a drain is present on the last site $\ket{N}$. The average transfer time is defined as~\cite{chahan1,chahan2,lloydfmo,lloydENAQT}
\begin{equation}
    \tau=\frac{\gamma_d}{\hbar} \int_0^\infty t \, |\Psi_N(t)|^2dt~,
    \label{eq1}
\end{equation}
where $\Psi_N(t)$ is the probability amplitude on the drain site   at time $t$, evolved under the effective Hamiltonian $H_{\rm eff}$~\cite{fmo2012,lev2017},
\begin{equation}
    (H_{\rm eff})_{k,l}= (H)_{k,l}-i\,\frac{\gamma_d}{2}\, \delta_{k,N}\delta_{l,N},
    \label{Hefftau}
\end{equation}
with $H$ given in  Eq.~\eqref{HAcp} and the non-Hermitian term representing the drain. 
A  rate equation can be derived, by assigning a drain frequency $1/\tau$ and a pumping frequency $\gamma_p/\hbar$, connecting  the chain  population $P_e$ to  the vacuum state $\ket{0}$ with population $P_0$:
\begin{align}
    &\frac{d P_0}{dt} = -\frac{\gamma_p}{\hbar}\, P_0 + \frac{1}{\tau}P_e, \nonumber \\
    &P_0 + P_e =1. \label{P0-2}
\end{align}
From the steady-state populations $P_e^{\rm SS}=\gamma_p/(\gamma_p+\hbar/\tau)$ we obtain the   current   $I=P_e^{\rm SS}/\tau $ and  its typical value,
\begin{equation}
    I^{\rm typ} = e^{\langle \ln I \rangle} \quad \mbox{with} \quad
     \langle \ln I \rangle \equiv \left \langle \ln \left( \frac{1}{\tau } \frac{\gamma_p}{\gamma_p+\frac{\hbar}{\tau }}\, \right) \right\rangle \,,
    \label{I-tau}
\end{equation}
%
where $\langle \cdots \rangle$ represents the average over disorder configurations. 

Another important figure of merit for the transport is the average variance $\langle \sigma^2 \rangle$ of the  excited states $|\alpha\rangle$,  defined as $\sigma^2 =[1/(N-1)] \sum_{\alpha=1}^{N-1} \sigma^2_{\alpha}$, where  $\sigma^2_\alpha \equiv \langle \alpha| {x^2}|\alpha \rangle -\langle \alpha |x  |\alpha\rangle^2$. This
  can be related to   the stationary variance obtained from the dynamical spreading of a  wave packet initially localized at the center of the chain; see SM. Moreover, in the SM, we also considered 
 another figure of merit for transport, i.e., the integrated transmission.
 Transport properties  revealed by the three different figures of merit are qualitatively the same.

{\it Results for  long-range systems.---}In Figs.~\ref{Fig:1}(a) and \ref{Fig:1}(b), $\hbar I^{\rm typ}/\Omega$, see Eq.~\eqref{I-tau}, and $\langle \sigma^2 \rangle$  are shown as a function of the normalized disorder strength $W/\Omega$ for a chain with $N=10^4$ sites. 
For small disorder both quantities decrease with $W$ exponentially, similarly to   the Anderson model ($\gamma=0$, blue curves). Counterintuitively, by increasing $W$, the transport efficiency at first increases   (DET regime),  until it reaches a  plateau, where the dependence on the disorder strength is extremely weak for several orders of magnitude of $W$ (DIT regime). The latter  persists approximately up to $W_{\rm gap}$.

Since the variance $\braket{\sigma^2}$ of the excited eigenstates, Fig.~\ref{Fig:1}(b), closely follows the behavior of the typical current $I^{\rm typ}$, Fig.~\ref{Fig:1}(a), we can try to understand the different transport regimes analyzing the average shape of the eigenfunctions $\braket{|\Psi|^2}$ of the excited states  as a function of the site basis $k$ for different disorder strengths $W$. 
  
\begin{figure*}[!ht]
    \centering 
    \includegraphics[width=\textwidth]{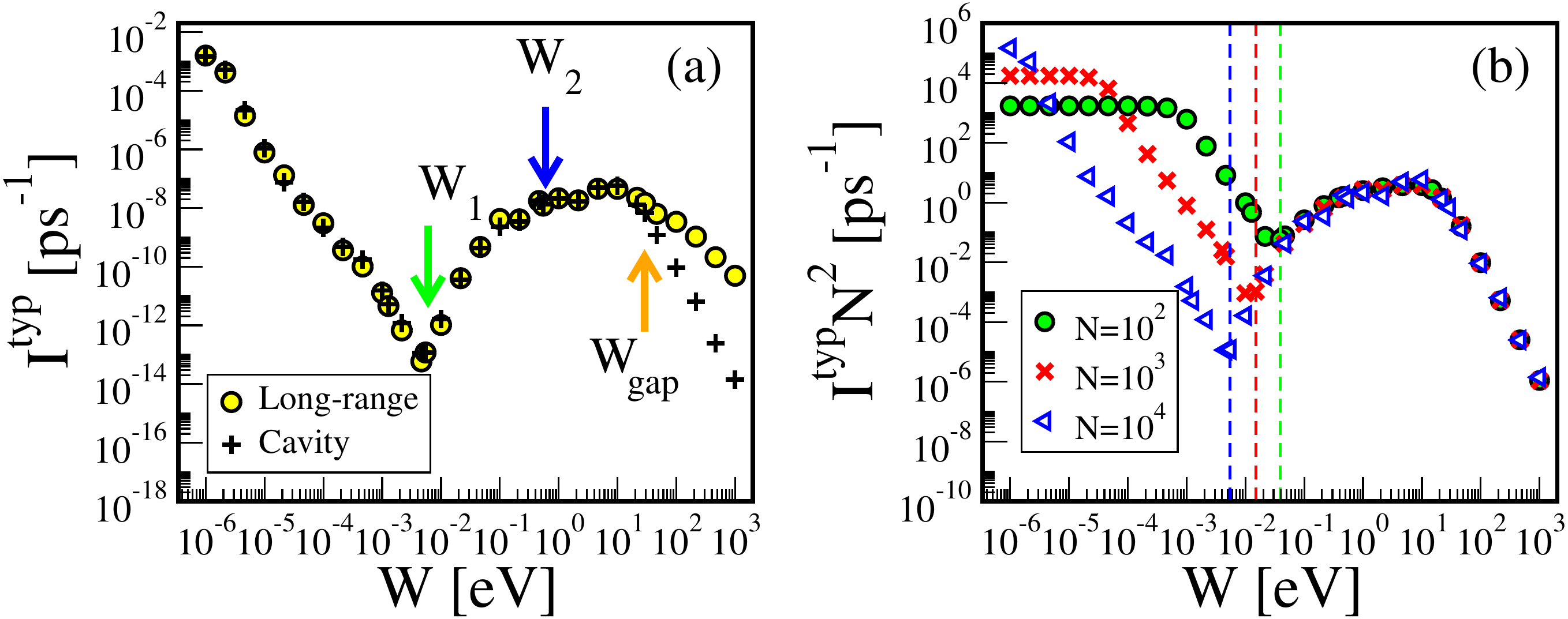}
    \caption{ (a) Typical  current $I^{\rm typ}$, Eq.~\eqref{I-tau}, versus the static disorder $W$. The results  for a linear chain in an optical cavity~Eq.~(\ref{Hcav}) (crosses) are compared with a long-range hopping model~Eq.~(\ref{HAcp})  (circles).
    Parameters for the linear chain in an optical cavity are $N=10^4$, $\Omega=0.0124$~eV, $\hbar\omega_c=2$~eV, $\mu \approx 36$~D, $g_c=3.188$~eV, $\gamma_p=\gamma_d=0.0124$~eV.
    The long-range hopping model has been obtained using the same $\Omega$ value and  setting $\gamma=2g_c/N$ in Eq.~\eqref{HAcp}.
    The number of disorder configurations $N_r$ is such that $N_r \times N =10^6$.  
    (b) Normalized typical  current $I^{\rm typ}N^2$ versus the static disorder $W$ for a  linear chain in an optical  cavity  for different $N$ values, as indicated in the legend. Vertical dashed lines represent the values of $W_1$ for different system sizes. Other parameters are the same as in (a).
   }
    \label{Fig4x}
\end{figure*}

Specifically, in the presence of long-range hopping~\cite{alberto,alberto2013,robincs}, in the gapped regime, the excited states  have a hybrid nature, with an exponentially localized peak, identical to the Anderson model peak,  and  extended flat tails, see  Figs.~\ref{Fig:1}(c) and \ref{Fig:1}(d), where the average shape of the eigenfunctions $\braket{|\Psi|^2}$ in the DET  and DIT regimes are shown.  Note that, while  in the DET regime the tails increase with the disorder strength $W$, they are independent of it in the DIT regime. Hybrid shapes of the eigenfuctions have been reported in other long-range interacting systems~\cite{prb2019algebraic,alberto}.  

An analytical expression for the disorder thresholds, separating the different transport regimes, can be found as follows.
When the probability of  the exponentially localized peak at the chain edges, $ \approx \exp(-N/2\xi)$, becomes equal to the average probability in  the tails (which scales as $1/N$; see SM), we have $\exp{(-N/2\xi)} \approx 1/N$. Recalling that $\xi \approx 105.2(\Omega/W)^2$, we get the disorder threshold $W_1$:
\begin{align}
     W_1 &\approx  \sqrt{\frac{210.4 \ln{N}}{N}} \,\Omega \,. 
    \label{W1}
\end{align}
%
For  $W>W_1$, the amplitude of the extended tails increases with the disorder strength $W$, see Fig.~\ref{Fig:1}(c), until the eigenfunction tails become independent of $W$; see Fig.~\ref{Fig:1}(d). The disorder threshold $W_2$ above which this happens  
 can be obtained by imposing that the probability on the closest sites to the peak is equal to the probability in the tails, $\exp(-1/2\xi) =1/N$, so that
\begin{align}
      W_2 &\approx \sqrt{210.4 \ln{N}}\Omega\,.
    \label{W2}
\end{align}
The validity of the predicted scaling of the different transport regimes with $N$ and $\gamma$ is discussed below and also  in the SM.  

One might think that these interesting transport regimes originate from  the coupling induced by disorder between the unperturbed excited states and the extended unperturbed ground state. Even if this coupling exists, it is not the main reason for  the DET and DIT regimes.  Indeed, a  semianalytical perturbative expression for the eigenstates in the gapped regime allows us to compute all the relevant observables, see orange dots in Figs.~\ref{Fig:1}(a) and \ref{Fig:1}(b), completely neglecting the coupling mediated by disorder between the unperturbed excited states and the extended unperturbed ground state; see details in SM.
This indicates that the DET and DIT regimes have their origin in  the existence of an extended ground state which, by imposing an orthogonality condition on all the excited states, generates their extended tails. 

{\it Applications to molecular chains in optical cavities.---}
Here we show that  a chain of emitters in a cavity~\cite{vidal,pupillo-old,pupillo-new} can be described in terms of an effective long-range hopping model arising from the coupling of the emitters with the cavity mode. This implies that our results are relevant for a vast variety of other systems such as Rydberg atoms, polar molecules, and molecular chains~\cite{pupillo-old}.  
%

In the following we focus on  the case in which the emitters are molecules. This is particularly interesting due to the large coupling (comparable with $k_BT$ with $T=300$~K) between the molecules. Nevertheless the same discussion can be applied to any other kind of emitters.
For  a molecular chain,  at resonance with a cavity mode~\cite{pupillo-old,pupillo-new} the Hamiltonian is given by
\begin{equation}
    H_{cav} = H_0 + g \sum_{j=1}^N ( \ket{j}\bra{c} + \ket{c}\bra{j} ) \, ,
    \label{Hcav}
\end{equation}
where $H_0$ is defined in Eq.~\eqref{Ho} and $\ket{c}$ represents a single excitation in the cavity mode (with no excitation in the chain). 
The coupling  $g$ of the emitters with the resonant optical mode is given by~\cite{scullybook}
\begin{equation}
g = \sqrt{ \frac{2\pi \mu^2 \hbar\omega_c} {V_c}} \, ,
\label{g}
\end{equation}
where $\mu$ is the molecular transition dipole, $\omega_c$ is the cavity mode frequency and  $V_c$ is the cavity mode volume. 

Since the coupling to the cavity mode is the same for all molecules, it is possible to show~\cite{vidal,pupillo-old} that only the fully symmetric state
$\ket{d}$ in the chain is coupled to the cavity mode with a collective coupling strength $g_c=\sqrt{N} g$. This coupling  induces two polaritonic states, $\ket {p_{\pm}} = 1/ \sqrt{2} \left(\ket{d} \pm \ket{c} \right )$, with an energy splitting of $2 g_c$, while the other $N-1$ states with a bandwidth $4 \Omega$, in the absence of disorder, 
are decoupled from the cavity mode. In the strong coupling regime, $g_c \gg \Omega$, one of the polaritonic states will become the ground state of the system and it will be gapped from the excited states by an energy $\approx g_c$. 
By imposing  
\begin{equation}
N \gamma_{\rm eff}/2= g_c,
\label{geff}
\end{equation}
we determine the effective long-range coupling $\gamma_{\rm eff}$ which would produce the same energy gap in the absence of disorder; see SM for details.

Since the coupling $g$ is inversely proportional to  $V_c$, see Eq.~(\ref{g}), which typically scales like $N$, 
in the following we consider a fixed collective coupling $g_c=\sqrt{N}g \approx 3.2$~eV~\cite{gambino2014,ultrastrong2013}, which 
corresponds to a cavity mode volume $V_c= 10^4$~nm$^3$~\cite{microcavity} for a molecular chain of $N=10^4$ with $\mu \approx 36$~D~\cite{expcavity}.

In Fig.~\ref{Fig4x}(a) we plot the typical current $I^{\rm typ}$ versus the disorder strength $W$ for a chain of $10^4$ molecules in an optical cavity (crosses). Interestingly, this current is reproduced extremely well by the current obtained with the effective long-range coupling~Eq.~(\ref{geff}) (circles) for $W<W_{\rm gap}$. For $W>W_{\rm gap}$ both polaritonic states mix with all the other states and the differences between the long-range model and the chain in the cavity model emerge. 
In Fig.~\ref{Fig4x}(b) the typical (normalized) current $I^{\rm typ} N^2$ for the cavity model, Eq.~(\ref{Hcav}), is shown for different chain sizes $N$.  Note that $I^{\rm typ} \propto 1/N^2$ for $W>W_1$, instead of decreasing exponentially with $N$, as for the localized regime in the absence of long-range hopping.   

{\it Conclusions.---}Controlling the detrimental effects of disorder at the nanoscale is one of the main challenges in achieving efficient energy transport. Here we have shown that long-range hopping can lead to a disorder-enhanced and a disorder-independent transport regime, extending over several orders of magnitude of disorder strength. 
Our results could be tested in several systems where long-range hopping is present, such as molecular aggregates~\cite{fmo2012}, ion traps~\cite{iontraps}, and cold atomic clouds~\cite{robin2017}. 
Remarkably, we have also shown that  a system of emitters coupled to a cavity mode  can be mapped  to a long-range hopping system.  This makes our results applicable to a vast variety of other physical systems, such as molecular chains in optical cavities, Rydberg atoms, and polar molecules~\cite{pupillo-old}; see SM for realistic parameters. 
Typically, for molecular chains in optical cavities $\Omega \approx 0.03$~eV, $N \approx 10^5$, and $g_c \approx 1$~eV~\cite{pupillo-old}, so that $W_1 \approx 5 \times 10^{-3}$~eV, $W_2 \approx 1.5$~eV, and $W_{\rm gap} \approx g_c \ln N \approx 11.5$~eV. Since natural disorder typically ranges from $1$ to $10$~$\Omega$, we can easily reach the DET regime, with currents in the measurable range of tens of nanoampere~\cite{expcavity}.
In other experimental setups, such as ion traps, the spreading of an initially localized excitation in the middle of the chain would provide the best way to access both the DET and DIT regime.
Indeed,
the stationary variance of the excitation, obtained from the spreading of a localized wave packet, is well described by the average variance of the eigenstates shown in Fig.~\ref{Fig:1}(b); see SM for details.
In perspective, it would be interesting to analyze the effect of thermal noise on transport in long-range systems. 

We acknowledge useful discussions with Thomas Botzung, Jérôme Dubail, Robin Kaiser, David Hagenmüller, Guido Pupillo, Johannes Schachenmayer and Stefan Schütz. F.B. and F.M. acknowledge support by the Iniziativa Specifica INFN-DynSysMath.  This work has been financially supported by the Catholic University of Sacred Heart within the program of promotion and diffusion of scientific research. Research has also been financially supported by Ministero dell’Istruzione, dell’Universit\`a e della Ricerca within the project PRIN 20172H2SC4. G.L.C. acknowledges the funding of CONACyT Ciencia Basica Project No.~A1-S-22706.


%

\end{document}


\title{Supplementary Material -- Disorder-Enhanced and Disorder-Independent Transport with long-range hopping: application to molecular chains in optical cavities}

\author{Nahum C. Ch\'avez}
\affiliation{Dipartimento di Matematica e Fisica and Interdisciplinary Laboratories for Advanced Materials Physics, Universit\`a Cattolica, via Musei 41, 25121 Brescia, Italy}
\affiliation{Benem\'erita Universidad Aut\'onoma de Puebla, Apartado Postal J-48, Instituto de F\'isica,  72570, Mexico}

\author{Francesco Mattiotti}
\affiliation{Dipartimento di Matematica e Fisica and Interdisciplinary Laboratories for Advanced Materials Physics, Universit\`a Cattolica, via Musei 41, 25121 Brescia, Italy}
\affiliation{Istituto Nazionale di Fisica Nucleare,  Sezione di Pavia, via Bassi 6, I-27100, Pavia, Italy}
\affiliation{Department of Physics, University of Notre Dame, Notre Dame, IN 46556, USA}

\author{J. A. M\'endez-Berm\'udez}
\affiliation{Benem\'erita Universidad Aut\'onoma de Puebla, Apartado Postal J-48, Instituto de F\'isica,  72570, Mexico}

\author{Fausto Borgonovi}
\affiliation{Dipartimento di Matematica e Fisica and Interdisciplinary Laboratories for Advanced Materials Physics, Universit\`a Cattolica, via Musei 41, 25121 Brescia, Italy}
\affiliation{Istituto Nazionale di Fisica Nucleare,  Sezione di Pavia, via Bassi 6, I-27100, Pavia, Italy}

\author{G. Luca Celardo}
\affiliation{Benem\'erita Universidad Aut\'onoma de Puebla, Apartado Postal J-48, Instituto de F\'isica,  72570, Mexico}

\maketitle

\tableofcontents

\setcounter{equation}{0} 
\setcounter{figure}{0} 
\renewcommand{\theequation}{S\arabic{equation}} 
\renewcommand{\thefigure}{S\arabic{figure}} 

\section{Realistic Parameters for different systems}
In the Main Text we have shown that long-range hopping can lead to a disorder-enhanced and a disorder-independent transport (DET and DIT) regimes, extending over several orders of magnitude of disorder strength $W$. We have also shown
that a realistic system consisting of a linear chain of emitters in an optical cavity can be mapped  to a long-range hopping system. This makes our results applicable to a vast number of physical systems, such as molecular chains, Rydberg atoms, polar molecules and ion traps, to mention a few. Below we give some realistic parameters for the Hamiltionian in Eq.~(12) in the Main Text, with respect to different physical systems. 
Typically for molecular chains in optical cavities $\Omega \approx 0.03$~eV, $N \approx 10^5$ and $g_c \approx 1$~eV~\cite{pupillo-old} so that $W_1 \approx 5 \times 10^{-3}$~eV, $W_2 \approx 1.5$~eV and $W_{\rm GAP} \approx g_c \ln N \approx 11.5$~eV. Since natural disorder  typically ranges from $1-10$~$\Omega$   we can easily reach the DET regime.  
Moreover several other systems consisting of emitters  in a cavity could display the same transport properties predicted in the Main Text, such as Rydberg atoms~\cite{pupillo-old}  for which we have $\Omega \approx 80~\mbox{kHz}$ and $\gamma_{\rm eff} \approx 3~\mbox{kHz}$, polar molecules where $\Omega \approx 50$~Hz and $g \gg \Omega$ or ion traps~\cite{iontraps} where $\Omega =0$ and $\gamma \approx 400~\mbox{Hz}$.

\section{Energy Gap and Long Range Interaction}

In the Main Text we have shown that adding long-range hopping to one-dimensional (1D) disordered quantum wires leads to a  finite energy gap $\Delta$ between the ground state and the excited states which protects the system from disorder~\cite{nahum}.

\begin{figure*}[t!]
    \centering
    \includegraphics[width=0.85\textwidth]{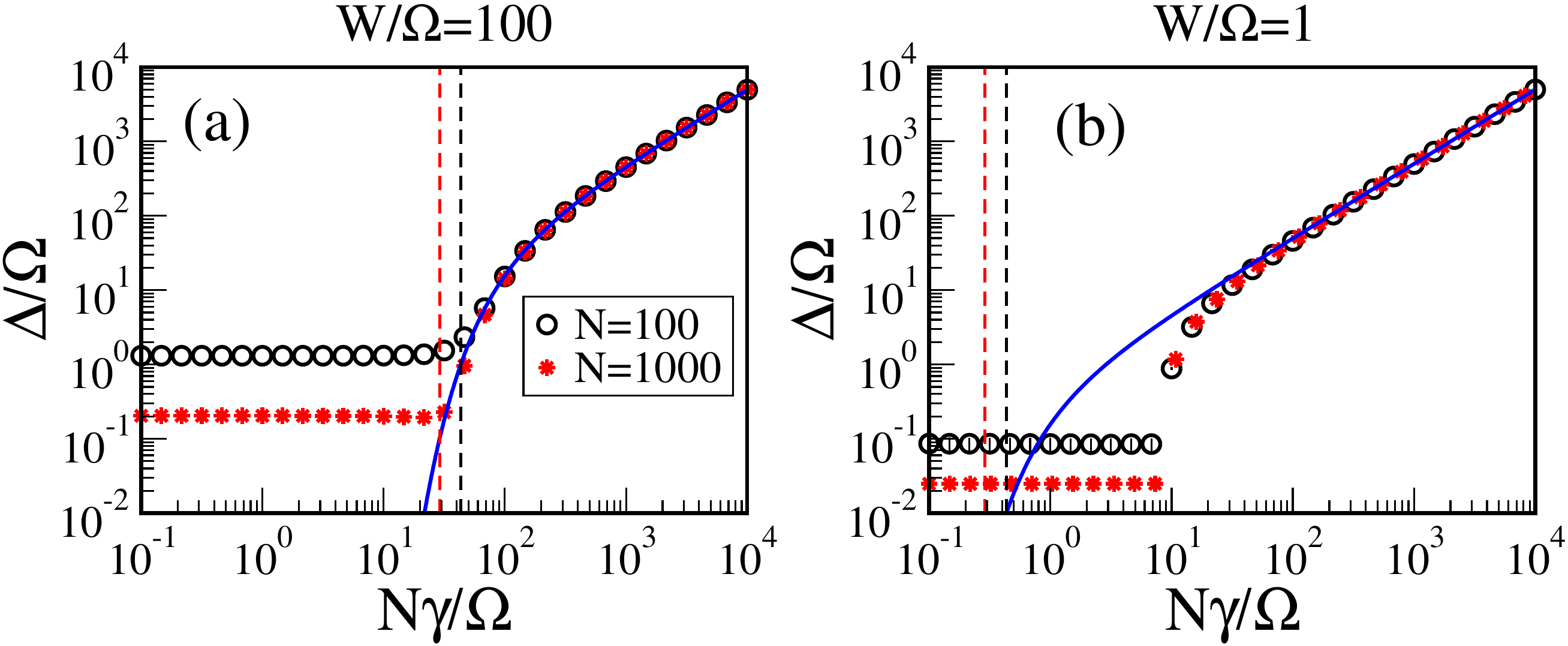}
    \vspace{-3mm}
    \caption{Energy gap $\Delta$ divided by the nearest-neighbor coupling $\Omega$ as a function of $N \gamma/\Omega$ for the  disorder strengths (a) $W/\Omega=100$ and (b) $W/\Omega=1$. Symbols are given by Eq.~\eqref{Eq:Dh}, while the continuous blue curves show the analytical estimate of Eq.~\eqref{gaph}. The dashed vertical lines indicate the critical coupling strength $\gamma_{\rm GAP}$ from Eq.~\eqref{gammagap}. Here, $N=\{100,1000\}$ and $N_r=100$ disorder configurations were used.}
    \label{Gap:And}
\end{figure*}

In Fig.~\ref{Gap:And} we plot the energy gap $\Delta$ divided by the nearest-neighbor coupling $\Omega$ as a function of the coupling strength $\gamma$ (multiplied by $N/\Omega$) for two values of wire size $N$ and two disorders strengths: $W/\Omega=100$, Fig.~\ref{Gap:And}(a), and $W/\Omega=1$, Fig.~\ref{Gap:And}(b). Here, we compute $\Delta$ as~\cite{nahum}
\begin{equation}
  \Delta =\max_i \left\{ \min_{j \ne i}\left[ \text{dist} \left(E_i, E_j\right) \right] \right\} \, ,
    \label{Eq:Dh}
\end{equation}
where $\{ E_i\}$ are the eigenvalues of the Hamiltonian
\begin{equation}
H= \sum_{j=1}^N \epsilon_j \ket{j}\bra{j} + \Omega \sum_{j=1}^{N-1} ( \ket{j}\bra{j+1}+\ket{j+1}\bra{j}) -\frac{\gamma}{2} \sum_{i \ne j} | i \rangle \langle j| \, ,
\label{HAcp}
\end{equation}
see Eqs.~(1,2) in the Main Text, and $\text{dist} \left(E_i, E_j\right)=|E_i-E_j|$.

Note that in Fig.~\ref{Gap:And} we report the average value of $\Delta$ over  disorder configurations. 
From Fig.~\ref{Gap:And} we observe that
below a critical coupling strength $\gamma_{\rm GAP}$, $\Delta$ remains constant as a function of $N\gamma$ but decreases for increasing $N$, while above $\gamma_{\rm GAP}$, $\Delta$ is an  increasing function of $N\gamma$ but it is independent of $N$. A good approximation for $\gamma_{\rm GAP}$ can be obtained from the expression for the disorder threshold derived in Ref.~\cite{nahum}
\begin{equation}\label{WGAP}
    W_{\rm GAP}=\frac{\gamma}{2} N \ln{N}
\end{equation}
as 
\begin{equation}
\gamma_{_{\rm GAP}}= \frac{2 W}{N\ln{N}} \ .
\label{gammagap}
\end{equation}
It is relevant to stress that Eq.~\eqref{gammagap} was obtained in Ref.~\cite{nahum} for the Picket-Fence model with all-to-all coupling. The fact that Eq.~\eqref{gammagap} works very well in estimating $\gamma_{\rm GAP}$ at large disorder strengths $W$, see the vertical dashed lines in Fig.~\ref{Gap:And}(a), allows us to anticipate that the expression 
\begin{equation}
  \Delta=\frac{W}{e^{2W/N\gamma}-1} \ ,
\label{gaph}
\end{equation}
also obtained in Ref.~\cite{nahum}, may describe well $\Delta$ above $\gamma_{\rm GAP}$ for the 1D Anderson model subject to long-range hopping. Indeed, the good correspondence between Eq.~\eqref{gaph} (blue-full curve) and the numerically obtained $\Delta$ (symbols) is clearly shown in Fig.~\ref{Gap:And}(a).

Finally, it is important to add that even if Eq.~\eqref{gammagap} does not provide good predictions for $\gamma_{\rm GAP}$ for small disorder strengths $W/\Omega$, see Fig.~\ref{Gap:And}(b), the analytical expression of Eq.~\eqref{gaph} still describes well $\Delta$ for large $N\gamma$.


\section{Transmission}
\label{secTrans}

In the Main Text we analyze two figures of merit to characterize the transport efficiency of 1D disordered quantum wires in  presence of  long-range hopping: the typical current $I^{\rm typ}$ and the average variance $\braket{\sigma^2}$ of the excited eigenstates. Here we report a third figure of merit: the transmission $T$, which is widely used in transport studies of low-dimensional disordered quantum systems. We note that according to the experimental set-up either the current as computed in the Main Text or the transmission as discussed here are the relevant figures of merit for transport. Indeed, the integrated transmission $T_{\rm int}$ considered here is  relevant in charge transport in presence of a large bias, or when dealing with transmission of an energy broad-band incoming beam. 

\begin{figure*}[t!]
	\centering
	\includegraphics[width=0.75\textwidth]{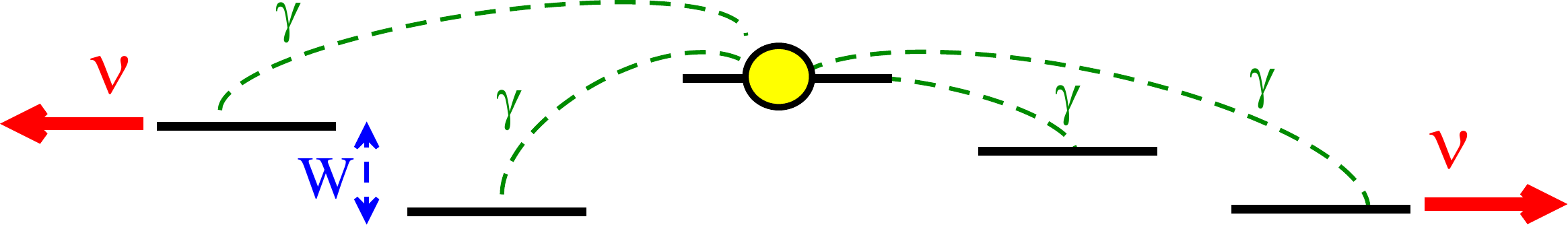}
	\vspace{-2mm}
	\caption{Pictorial representation of the scattering setup. The black lines represent the Anderson model spectrum with a random level spacing $W$ while the yellow circle is an excitation which can hop to all the other states with an amplitude $\gamma$ (green dashed lines). The excitation can then be collected by the leads at the edge sites with an amplitude $\nu$.
	}
    \label{Fig:setup}
\end{figure*}

The transmission through the 1D chain  can be studied by turning the setup of Fig.~1(a) of the Main Text into a scattering setup. To this end we couple the first and the last sites of the chain (i.e.~sites 1 and $N$) to two different perfect leads with coupling strength $\nu$, so that the components of the {\it effective} Hamiltonian (i.e.~the Hamiltonian of the scattering setup) read~\cite{Luca09} 
\begin{equation}
  (H_{\rm eff})_{k,l} = (H)_{k,l}-\frac{i}{2} \nu (\delta_{k,1}\delta_{l,1}+\delta_{k,N}\delta_{l,N}) \, ,
\label{H-T}
\end{equation}
where $H$ is given in Eq.~\eqref{HAcp}. A pictorial representation of the scattering setup is shown in Fig.~\ref{Fig:setup}, where an excitation which can hop among the chain sites is shown as the yellow circle.

In general, the transmission $T^{a,b}(E)$ from channel $a$ to channel $b$ can be determined by~\cite{Luca09}
\begin{equation}
T^{a,b}(E)= | Z^{a,b}(E) |^2 \ ,
\label{Eq-T}
\end{equation}
where 
\begin{equation}
Z^{a,b}(E)=\sum_{i,j=1}^N A^a_i \left(\frac{1}{E-H_{\rm eff}}\right)_{i,j} A^b_j
\label{Eq-ZT}
\end{equation}
is the transmission amplitude, $H_{\rm eff}$ is the effective non-Hermitian Hamiltonian in Eq.~\eqref{H-T} and $A_i^a$ are the decay amplitudes from the discrete internal states $i$ to the external states $a$.
Alternatively, we can also write $T^{a,b}(E)$ by diagonalizing the Hamiltonian $H_{\rm eff}$. The eigenfunctions of $H_{\rm eff}$, $\ket{r}$ and $\bra{ \tilde{r}}$, form a bi-orthogonal complete set,
\begin{equation}
H_{\rm eff}\ket{r}=\mathcal{E}_r \ket{r}~, \qquad \bra{ \tilde{r}}  H_{\rm eff}=\bra{\tilde{r}} \mathcal{E}_r,
\label{Eq-Heff}
\end{equation}
and its eigenenergies are complex numbers with the form
\begin{equation}
\mathcal{E}_r=E_r-\frac{i}{2}\Gamma _r \, ,
\label{Eq-Eg}
\end{equation}
corresponding to resonances centered at the energy $E_r$ with decay widths $\Gamma_r$. The decay amplitudes $A_i^a$ are thus transformed according to
\begin{equation}
\mathcal{A}^a_r=\sum_i A^a_i \braket{ i | r }, \; \; \; \mathcal{\tilde{A}}^b_r=\sum_j  \braket{\tilde{r} | j } A^b_j \; \; ,
\label{Eq-AHeff}
\end{equation}
and the transmission amplitude $Z^{a,b}(E)$ is then given by
\begin{equation}
Z^{a,b}(E)=\sum_{r=1}^N \mathcal{A}^a_r\frac{1}{E-\mathcal{E}_r}\mathcal{\tilde{A}}^b_r \, .
\label{Eq-ZHeff}
\end{equation}
Note that the complex eigenvalues $\mathcal{E}$ of $H_{\rm eff}$ coincide with the poles of the transition amplitude $Z(E)$.

Since the excitation is collected by the leads at the edges of the chain, the amplitudes in Eq.~\eqref{Eq-AHeff} are $\mathcal{A}^a_r=\sqrt{\nu} \Psi^1_r$ and $\mathcal{\tilde{A}}^b_r=\sqrt{\nu} \Psi^{N \ast}_r $,  where $\Psi^{1,N}_r= \langle 1,N|r\rangle$ is the amplitude of the eigenfunction $\ket{r}$  of the effective Hamiltonian on sites $1,N$. 
Moreover, the transmission amplitude $Z^{a,b}(E)$ in Eq.~\eqref{Eq-ZHeff} becomes 
\begin{equation}
Z^{a,b}(E)=\nu \sum_{r=1}^N  \frac{\Psi^1_r \Psi^{N \ast}_r}{E-\mathcal{E}_r} .
\label{Eq-Zint}
\end{equation}
Since the conjugate of the transmission amplitude $Z^{a,b}(E)$ is
\begin{equation}
Z^{a,b}(E)^\ast=\nu \sum_{k=1}^N  \frac{\Psi^{1 \ast}_k \Psi^N_k}{E-\mathcal{E}_k^\ast} ,
\label{Eq-Zintconj}
\end{equation}
the transmission $T^{a,b}(E)=| Z^{a,b}(E) |^2=Z^{a,b}(E) Z^{a,b}(E)^\ast$ is written as
\begin{equation}
T^{a,b}(E)= \nu^2 \sum_{r=1}^N \sum_{k=1}^N \frac{\Psi^1_r \Psi^{1 \ast}_k \Psi^N_k \Psi^{N \ast}_r }{(E-\mathcal{E}_r)(E-\mathcal{E}_k^\ast )}  \, .
\label{Eq-Tint0}
\end{equation}

Now, let us integrate Eq.~\eqref{Eq-Tint0} over all the energies, i.e.~$T_{\rm int}=\int_{-\infty}^{\infty} dE \, T^{a,b}(E)$,
so we get the expression
\begin{equation}
T_{\rm int} = \displaystyle \nu^2 \sum_{r=1}^N \sum_{k=1}^N \Psi^1_r \Psi^{1 \ast}_k \Psi^N_k \Psi^{N \ast}_r \int_{-\infty}^{\infty}  \frac{dE}{(E-\mathcal{E}_r)(E-\mathcal{E}_k^\ast )} = \displaystyle 2 \pi \nu^2 \sum_{r=1}^N \sum_{k=1}^N \frac{ \Psi^1_r \Psi^{1 \ast}_k \Psi^N_k \Psi^{N \ast}_r }{(\Gamma_r+\Gamma_k)/2-i(E_k-E_r)} \, .
\label{Eq-Tint}
\end{equation}
Let us note that Eq.~\eqref{Eq-Tint} is exact and it depends only on the amplitudes of the eigenfunctions at the edges of the chain $\Psi^{1,N}_r$ and the complex eigenvalues $\mathcal{E}_r$ of $H_{\rm eff}$. 
The integrated transmission $T_{\rm int}$ represents the overall transmission over a wide spectral energy band and, for instance, is relevant for analyzing the transport under a large applied voltage:
the shape of the current-voltage characteristic can sometimes be significantly different depending on the potential profile or the voltage drop. This is important in determining the maximum current of a transistor~\cite{datta}.

\begin{figure*}[t!]
  \centering
  \includegraphics[width=0.85\textwidth]{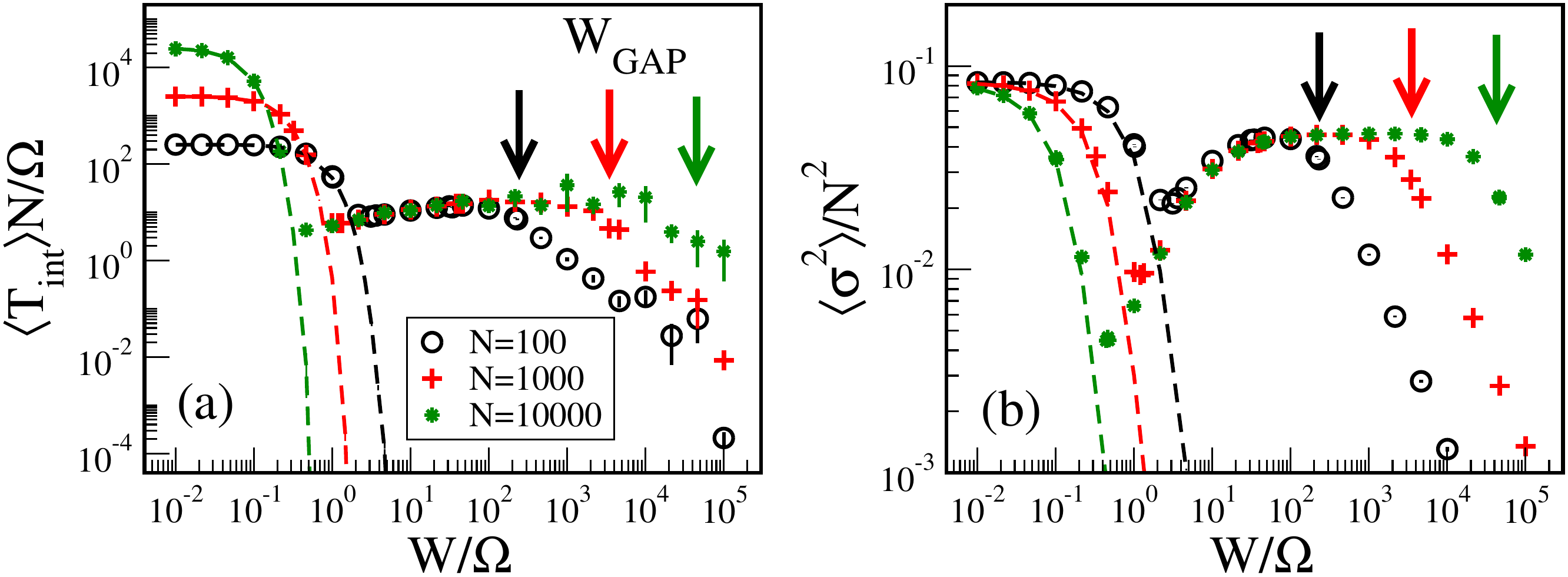}
  \vspace{-3mm}
  \caption{(a) Rescaled average integrated transmission $\braket{T_{\rm int}} N/\Omega $ 
  and (b)
  rescaled average variance $\braket{\sigma^2}/N^2$
   as a function of the rescaled  disorder strength $W/\Omega$ for different system sizes $N$ as indicated in the legend.
  Here, we choose the coupling strength to the leads (a) $\nu=\Omega$ and (b) $\nu=0$,  $\gamma=\Omega$ and the disorder configurations $N_r$ are such that $N_r \times N=10^5$.
  Arrows mark the critical disorder $W_{\rm GAP}$ for each system size $N$ according to Eq.~\eqref{WGAP}. The dashed (green, red and black) curves show the cases $\gamma=0$.}
  \label{Fig3}
\end{figure*}
\begin{figure*}[ht!]
  \centering
  \includegraphics[width=0.84\textwidth]{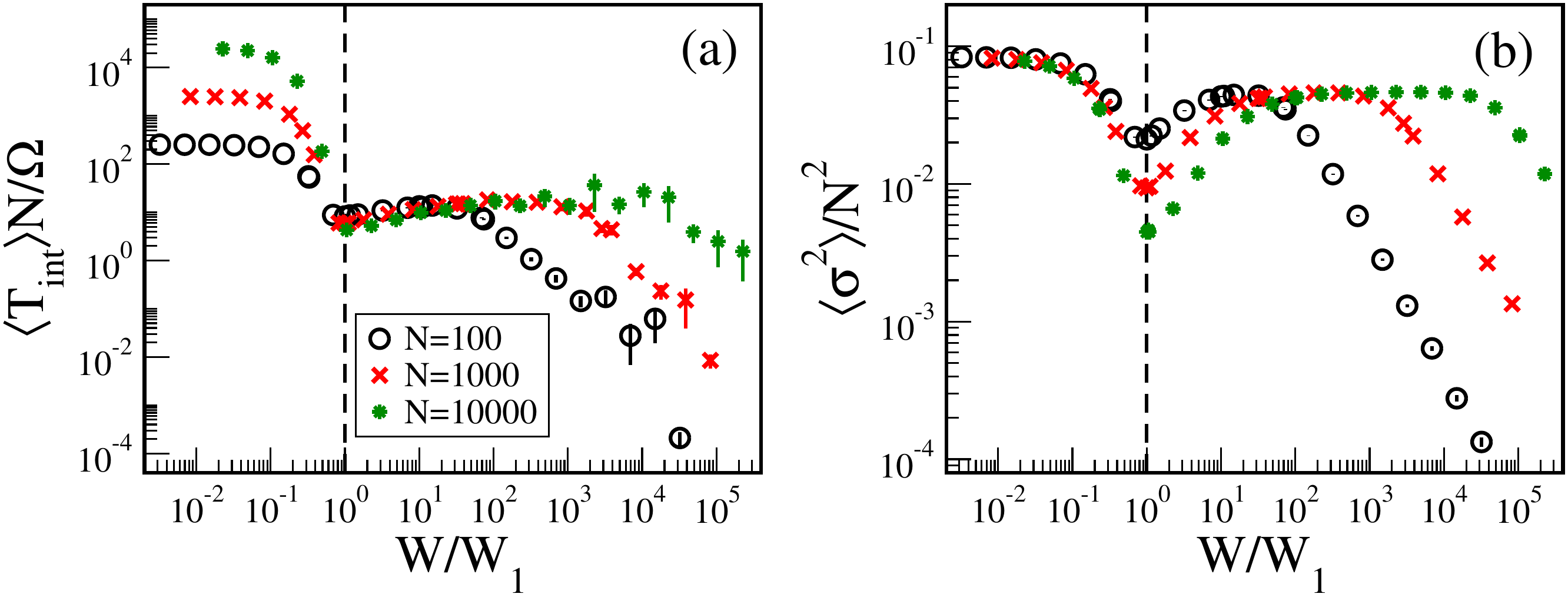} \\
  \includegraphics[width=0.84\textwidth]{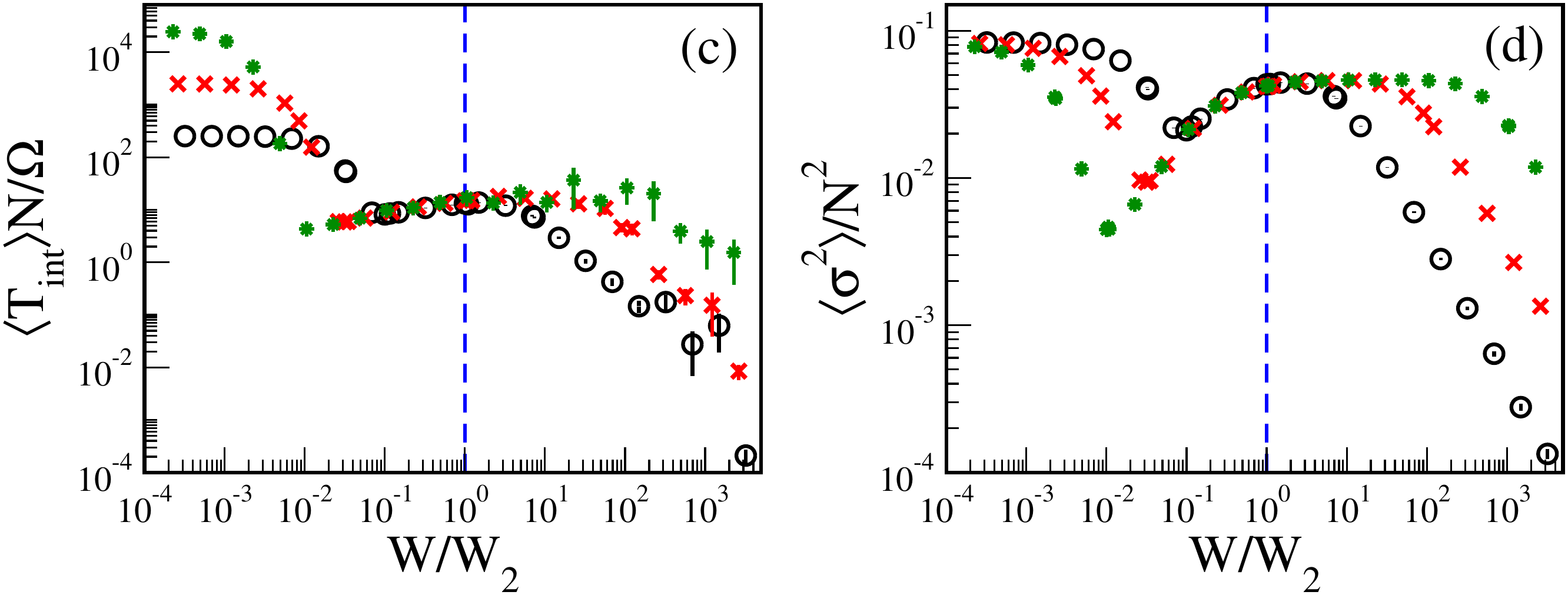}
  \vspace{-4mm}
  \caption{(a) Rescaled average integrated transmission $\braket{T_{\rm int}}\!N/\Omega$ 
  and (b) rescaled average variance $\braket{\sigma^2}/N^2$ as a function of the normalized static disorder $W/W_1$. (c) Rescaled average integrated transmission $\braket{T_{\rm int}}\!N/\Omega$ 
  and (d) rescaled average variance $\braket{\sigma^2}/N^2$ as a function of the normalized static disorder $W/W_2$. Vertical dashed lines indicate the critical disorders (a,b) $W_1$ and (c,d) $W_2$, which are computed from Eqs.~\eqref{Wcr_1} and~\eqref{W2_2}, respectively. Same parameters of Fig.~\ref{Fig3} were used.} 
  \label{Fig:13}
\end{figure*}


In Fig.~\ref{Fig3}(a) we present the average integrated transmission $\braket{T_{\rm int}}$, multiplied by $N/\Omega$, as a function of the normalized static disorder $W/\Omega$ for the coupling strength to the leads $\nu=\Omega$; three wire lengths are reported, i.e. $N=\{100,1000,10000\}$. As a reference, we also present the case of the 1D Anderson model without the long-range hopping, i.e.~the $\gamma=0$ case, see the dashed (green, red and black) curves. Interestingly the integrated transmission $T_{\rm int}$ in the DIT regime decays as $1/N$ in contrast to the Anderson model case where it decays exponentially with the system size $N$. Moreover, by comparing Fig.~\ref{Fig3}(a) with Fig.~2 in the Main Text, it becomes clear that all the features reported there for the typical current $I
^{\rm typ}$ and the average variance $\braket{\sigma^2}$ as a function of $W$ are also present in the average integrated transmission $\braket{T_{\rm int}}$. 
Moreover, in Figs.~\ref{Fig:13}(a) and~\ref{Fig:13}(c) we plot the curves of $\braket{T_{\rm int}}N/\Omega$ of Fig.~\ref{Fig3}(a) but now as a function of the static disorder $W$ normalized by $W_1$ and $W_2$, respectively, see Eqs.~(10,11) in the Main Text. With this we verify that the estimations for the critical disorders $W_1$ and $W_2$ (derived in the Main Text and in the following Section), as given in Eqs.~\eqref{Wcr_1} and~\eqref{W2_2}, respectively, work well for the average integrated transmission $\braket{T_{\rm int}}$.

In addition, for comparison purposes, in Figs.~\ref{Fig3}(b), \ref{Fig:13}(b) and~\ref{Fig:13}(d) we present $\braket{\sigma^2}/N^2$ for the same parameter values used in Figs.~\ref{Fig3}(a), \ref{Fig:13}(a) and~\ref{Fig:13}(c), respectively.
$\braket{\sigma^2}/N^2$  represents the normalized average variance of the excited eigenstates, defined as 
\[
 \sigma^2 =\dfrac{1}{N-1} \displaystyle\sum_{\alpha=1}^{N-1} \sigma^2_{\alpha} \quad \mbox{where} \quad   \sigma^2_\alpha \equiv \braket{x_\alpha^2}-\braket{x_\alpha}^2 \quad \mbox{and} \quad \braket{x_\alpha^2} = {\displaystyle\sum_i i^2 \,\,|\braket{i|\alpha}|^2}, \quad \braket{x_\alpha} = \displaystyle\sum_i i \,\,|\braket{i|\alpha}|^2 .
\]
A good correspondence in the behavior of the curves for the average integrated transmission $\braket{T_{\rm int}}$ and the average variance $\braket{\sigma^2}$, as a function of $W$, is clearly observed.

\begin{figure*}[t!]
  \centering
  \includegraphics[width=0.85\textwidth]{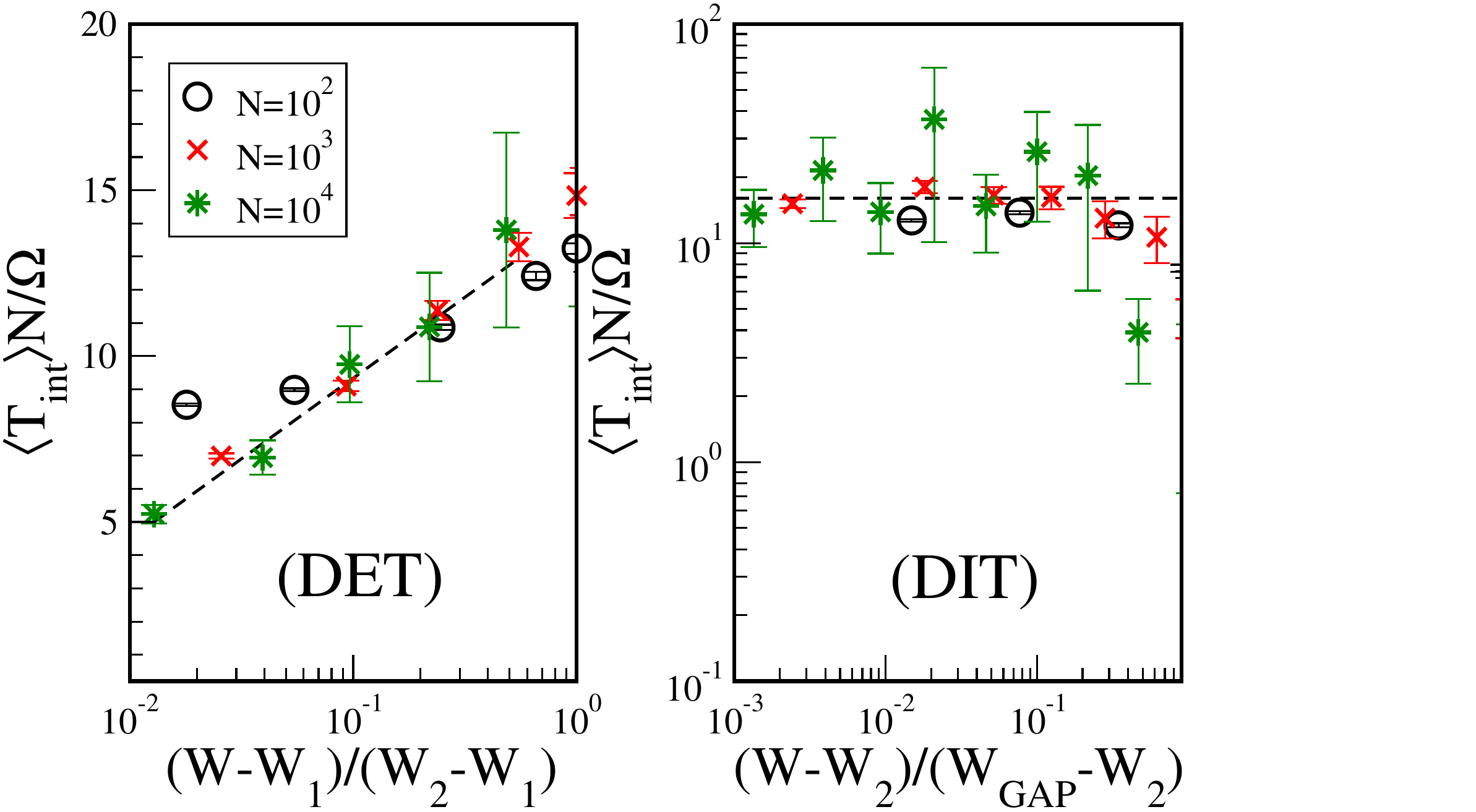}
  \vspace{-3mm}
  \caption{Left panel (DET): Rescaled average integrated transmission $\braket{T_{\rm int}} N/\Omega $  {\it vs.} the rescaled disorder strength $(W-W_1)/(W_2-W_1)$.
  The dashed line is a logarithmic fit of all the symbols drawn in order to show the increase of transmission. Right panel (DIT): Rescaled average integrated transmission $\braket{T_{\rm int}} N/\Omega $  {\it vs.} the rescaled disorder strength $(W-W_2)/(W_{\rm GAP}-W_2)$.
  The dashed line is a linear fit of all the symbols drawn in order to show that the transmission is approximately constant.
  The parameters are the same as in Fig.~\ref{Fig:13}. The error bars indicate one standard deviation.
    }
  \label{Fig13a}
\end{figure*}

In order to make the two different regimes more explicit, we rescale the disorder strength in the following way:
for the DET
regime
we put on the $x$-axis the variable 
$$W^\prime = (W-W_1)/(W_2-W_1)$$ so that 
$W^\prime(W_1)=0$ and $W^\prime(W_2)=1$.
In this way all data sets with different $N$ in the DET regime have 
$0<W^\prime<1$. As one can see in Fig.~\ref{Fig13a}(left panel), in the DET region all the points with different $N$ lie approximately on the same curve. To guide the eye we perform a logarithmic fit, see dashed line in the same panel.

In order to show that the transmission in the DIT regime is approximately constant we perform a similar change of variable, {\it i.e.}
$$W^{''} = (W-W_2)/(W_{\rm GAP}-W_2),$$ in such a way that in the DIT regime
$0<W^{''}<1$ for all different $N$.
As one can see in Fig.~\ref{Fig13a}(right panel), all data sets for different $N$ show that in this regime the transmission is approximately constant. To guide the eye we added a linear fit performed on all the points in this regime (see dashed line).

\section{Shape of Eigenfunctions}
\label{Rel-Eig-Trans}

\subsection{Relationship between the shape of eigenfunctions and transport properties}

The analysis of the shape of the eigenfunctions is essential to understand the transport properties of the system. 
Moreover, this analysis will allow us to explain the different transport regimes discussed above and to analytically estimate the different critical disorders discussed in the previous section.

The dependence of the shape of the eigenfunctions on the disorder strength $W$ in our model  is much richer than what we have in the Anderson model in absence of long-range hopping. 
Indeed in the 1D Anderson model, the eigenfunctions are always exponentially localized.

The situation is very different in presence of long-range hopping. 
For instance, the shape of eigenfunctions in 1D and 3D Anderson models with the addition of all-to-all non-Hermitian couplings have been already analyzed by some of the authors of this manuscript in Ref.~\cite{alberto} and the main results obtained about the shape of eigenfuctions are valid also for our case where the long-range coupling is Hermitian.  

Here we analyze
the shape of the eigenfunctions in the site basis for different disorder strengths $W$, fixed $\gamma$ and $N$, and no coupling to the leads, i.e.~$\nu=0$. In our numerical experiments, the average shape of the eigenfunctions $\braket{|\Psi|^2}$ has been obtained for each disorder configuration as follows:
\begin{enumerate}
\item We diagonalize the Hamiltonian given in Eq.~(1) of the Main Text and reproduced in Eq.~(\ref{HAcp}). 
\item We consider those eigenfunctions peaked within the $20\%$ of sites around the middle of the chain.
\item We shift the position of the selected eigenfunctions so that all maxima coincide.
\item We determine the average shape of the eigenfunctions $\braket{|\Psi|^2}$ by averaging their probability distributions. 
\end{enumerate}

\begin{figure*}[t]
    \centering
    \includegraphics[width=0.83\textwidth]{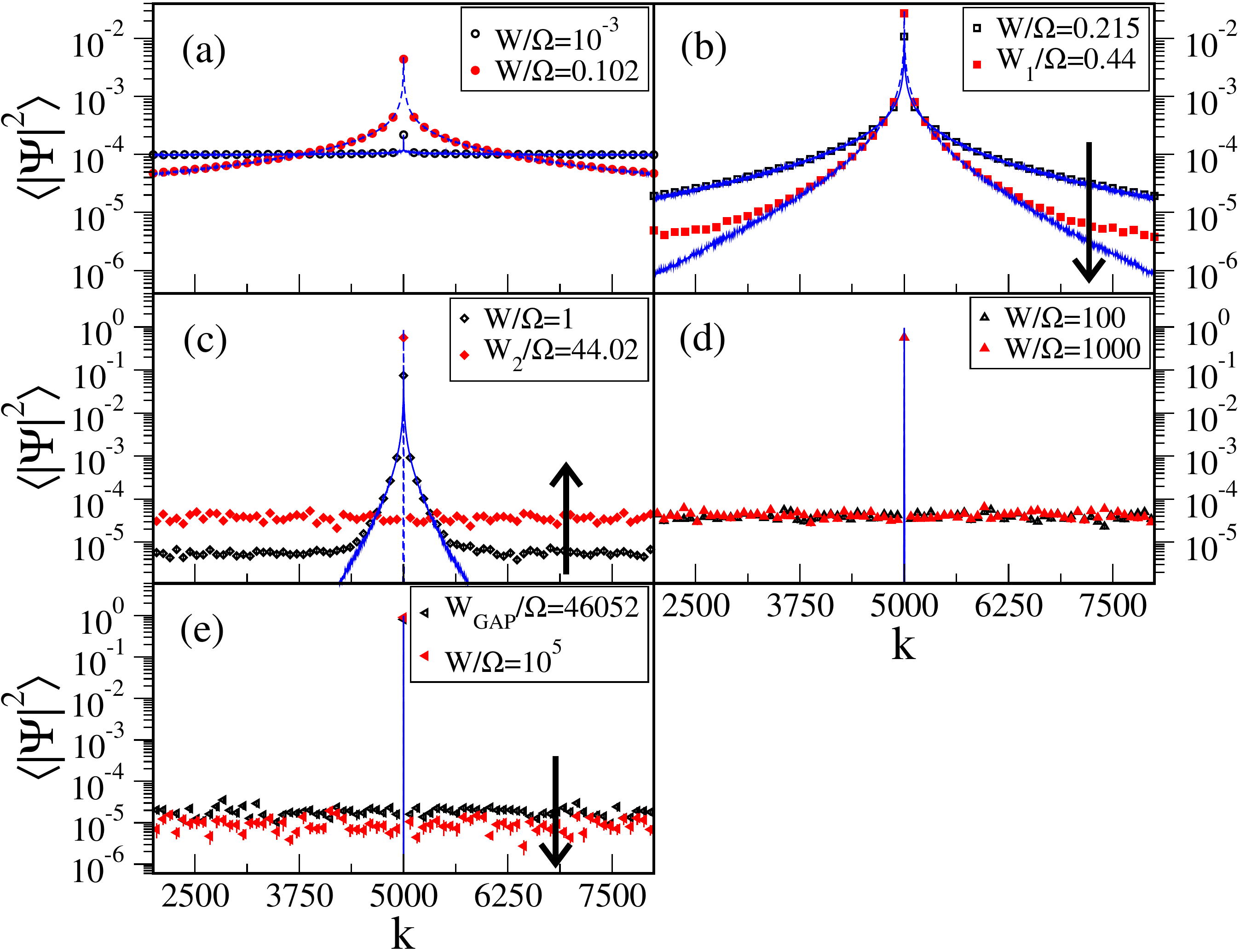}
    \vspace{-4mm}
   \caption{Average shape of the eigenfunctions $\braket{|\Psi|^2}$ in the site basis $k$. Different disorder regimes are shown
   in each panel: (a) $W\le 0.102~\Omega$, (b) $0.102~\Omega < W \le W_1$, (c) $W_1 < W \le W_2$, (d) $W_2 < W < W_{\rm GAP}$ and (e) $W \ge W_{\rm GAP}$. Here, $N=10^4$ and $\gamma=\Omega$. The averages are taken over $N_r=100$ disorder configurations. The continuous lines
   indicate the corresponding average shape of eigenfunctions $\braket{|\Psi|^2}$ for the case $\gamma=0$. Black arrows indicate increasing $W$.}
   \label{Fig:Shape-H}
\end{figure*}

In Fig.~\ref{Fig:Shape-H} we show the average  shape of the eigenfunctions $\braket{|\Psi|^2}$ in the site basis $k$ for the coupling strength $\gamma=\Omega$, system size $N=10^4$ and different disorder strengths $W$ as indicated in the legends. 
For all disorder strengths $W$, in each panel, we also show the average shape of the eigenfunctions $\braket{|\Psi|^2}$ for the corresponding Anderson model, i.e.~with $\gamma=0$.
By analyzing the average shape of the eigenfunctions $\braket{|\Psi|^2}$ we can identify the different disorder regimes which are relevant to understand the transport properties of the system, reported in the previous section:\\

\begin{enumerate}
\item ($W<W_1$) For very small disorder strength $W$,  the localization length $\xi$ of the eigenfunctions is larger than the systems size $N$, so that the eigenfunctions are delocalized. The shape of the eigenfunctions $\braket{|\Psi|^2}$ is similar to that of the Anderson model in the absence of long-range coupling, see Fig.~\ref{Fig:Shape-H}(a). As disorder increases, an exponential peak becomes visible, see Fig.~\ref{Fig:Shape-H}(b). In this regime the shape of the eigenfunctions $\braket{|\Psi|^2}$ is  similar to the shape of the eigenfunctions $\braket{|\Psi|^2}$ of the 1D Anderson model, up to the threshold  strength $W_1$, see Main Text.  
We can define this  disorder threshold $ W_1 $ taking into consideration that the eigenfuctions of the excited states have a hybrid character as discovered in Ref.~\cite{alberto}. Indeed they present an exponentially localized peak with the same localization length of the 1D Anderson model, and an extended tail which decreases with the system size as $1/N$. Thus, we can estimate the threshold disorder strength for which the eigenfunctions of the Anderson model with long-range hopping will differ from the eigenfuctions of the 1D Anderson model, by finding the disorder strength $W$
 for which the probability of  the exponentially localized peak at the chain edges becomes comparable to probability in  the extended tails. 
Considering that the extended tails in the gapped regime decrease as $1/N$  and considering that  the exponential peak at the chain edges is given by $\exp(-N/2\xi)$, we can determine $ W_1 $ by the following equation,
\begin{equation}
    \exp{\left(-\frac{N}{2\xi}\right)} \approx \frac{1}{N} \Rightarrow \quad -\frac{N}{2\xi} \approx -\ln{N} ,
    \label{wcr-tail}
\end{equation}
where the localization length $\xi$ for $E=0$~\cite{Izrailev98} is 
\begin{equation}
\xi(E=0)=105.2 \left(\frac{\Omega}{ \rm W}\right)^2 \, ,
\end{equation}
so that Eq.~\eqref{wcr-tail} becomes
\begin{align}
    -\frac{N}{210.4} \left(\frac{W_1}{\Omega}\right)^2 &\approx -\ln{N}\,, \nonumber \\
    \Rightarrow  W_1 &\approx  \sqrt{\frac{210.4 \ln{N}}{N}} \,\Omega \, .
    \label{Wcr_1}
\end{align}

\item ($W_1 \le W<W_2$) Above the disorder threshold $W_1$, the probability of the extended tails increases as the disorder strength $W$ increases, see black arrow in Fig.~\ref{Fig:Shape-H}(c), and the eigenfunctions change their shape: the probability in the extended tail increases and the peak becomes more localized as the disorder strength $W$ increases. 
The disorder threshold $W_2$ can be obtained by imposing the probability of the closest site to the peak to be equal to the probability in the extended tails.  
Considering that the tails in the gapped regime decrease as $1/N$, which is independent of the disorder strength $W$, and that the exponential peak on the closest site is given by $\exp(-1/2\xi)$, we can determine $ W_2 $ by the following equation,
\begin{equation}\label{wcr-peak}
    \exp{\left(-\frac{1}{2\xi}\right)} \approx \frac{1}{N} \Rightarrow \quad -\frac{1}{2\xi} \approx -\ln{N},
\end{equation}
so that Eq.~\eqref{wcr-peak} becomes
\begin{align}
    -\frac{1}{210.4} \left(\frac{W_2}{\Omega}\right)^2 &\approx -\ln{N}\,, \nonumber \\
    \Rightarrow  W_2 &\approx \sqrt{210.4 \ln{N}}\,\Omega\,.
    \label{W2_2}
\end{align}

\item ($W_2 \le W<W_{\rm GAP}$) Above the disorder threshold $W_2$, the eigenfunctions of the excited states are fully localized and  the amplitude of the extended tails is independent of the disorder strength $W$, see Fig.~\ref{Fig:Shape-H}(d).

\item ($W \ge W_{\rm GAP}$) Above the critical disorder $W_{\rm GAP}$, the eigenfunctions of the excited states are fully localized on one site with extended tails whose amplitude decreases as the disorder strength $W$ increases, see the vertical black arrow in Fig.~\ref{Fig:Shape-H}(e).

\end{enumerate}

The analysis of the average shape of the eigenfunctions $\braket{|\Psi|^2}$ indicates a strong correlation with the transport properties of the system. Specifically, we observe that the typical current $I^{\rm typ}$, the integrated transmission $T_{\rm int}$ and the variance $\sigma^2$ are independent of the  disorder strength $W$ in the same disorder range where the extended tails of the average shape of the eigenfunctions $\braket{|\Psi|^2}$ are independent, too.  
Thus, we can claim that the extended tails in the probability distribution of the eigenfunctions support the robustness of transport properties  in the gapped regime.

We stress that the analysis of the average shape of the eigenfunctions $\braket{|\Psi|^2}$ above also allowed us to determine the  disorder thresholds as a function of the model parameters, which define the different transport regimes. 

\begin{figure*}
  \centering
  \includegraphics[width=0.4\textwidth]{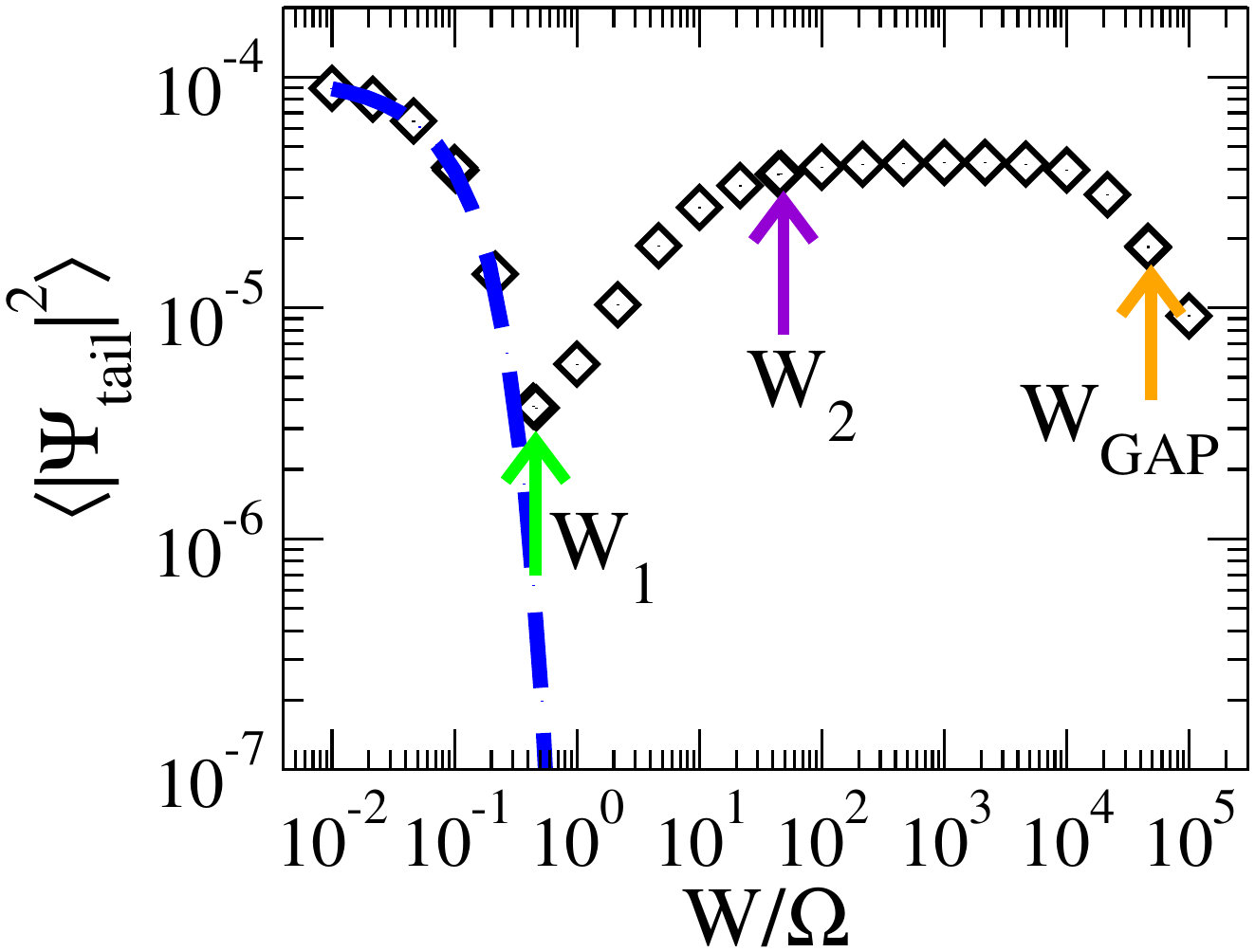}
  \caption{Average eigenfunction amplitude on the extended tails of the hybrid states, $\braket{|\Psi_{\rm tail}|^2}$, {\it vs.} the normalized static disorder $W/\Omega$ for the coupling strength $\gamma=\Omega$.
  Here, $N=10^4$. The average is performed over $N_r=10$ disorder configurations. The blue dashed line corresponds to the case $\gamma=0$. The arrows indicate the different  disorders thresholds given by Eqs.~\eqref{WGAP},~\eqref{Wcr_1} and~\eqref{W2_2}.  }
  \label{Fig:8}
\end{figure*}

It is relevant to notice that the panorama observed through the three figures of merit we used to characterize the transport efficiency of 1D disordered quantum wires in the presence of long-range hopping (i.e.~the typical current $I^{\rm typ}$, the average variance $\braket{\sigma^2}$ of the corresponding eigenstates and the integrated transmission $T_{\rm int}$) can be effectively reproduced from the average eigenfunction amplitude of the hybrid states on the extended tails, $\braket{|\Psi_{\rm tail}|^2}$. This is shown in Fig.~\ref{Fig:8}, where we plot $\braket{|\Psi_{\rm tail}|^2}$ as a function of the disorder strength $W$ for hybrid states at the coupling strength $\gamma=\Omega$. $\braket{|\Psi_{\rm tail}|^2}$ is computed from a triple average: over the components of the hybrid eigenfunction excluding the exponentially localized peak, over all hybrid eigenfunctions of a disorder wire and over an ensemble of disorder configurations. Indeed, the curve $\braket{|\Psi_{\rm tail}|^2}$ {\it vs.}~$W$ of Fig.~\ref{Fig:8} clearly displays all the characteristics of the transport properties, including the disorder-independent regime $W_2 \le W<W_{\rm GAP}$. 



We also want to remark that the hybrid states plateau has nontrivial statistical properties: the average value of the probability in the plateau goes like $1/N$, while the typical probability value goes like $1/N^2$. By typical probability we mean $\exp\braket{\ln |\Psi_{\rm tail}|^2}$. These two scalings are shown in Fig.~\ref{Psi} from the stationary probability distribution obtained by evolving a wave packet initially localized at the center of a linear chain in the DIT regime.

\subsection{Perturbative approach to the shape of eigenfuctions}
\label{appPert}




Since we demonstrated that the shape of the eigenfunctions allows to understand the transport properties of the system, here we intend to derive a perturbative expression for the shape of the excited eigenfunctions in the gapped regime. 
For this purpose, we rewrite Eq.~\eqref{HAcp} in its matricial form as follows
\begin{equation}
H = H_0 - \frac{\gamma}{2} Q + \frac{\gamma}{2} I \, ,
\label{Hfull}
\end{equation}
where $Q$ is a full matrix with components $1$ and $I$ is the identity matrix. 
The $Q$ matrix can be easily diagonalized. It has only two different eigenvalues.
The first eigenvalue, $-\gamma N/2$, corresponding to the lowest energy state, is a fully non degenerated extended state,
\begin{equation}
\ket{d}=\dfrac{1}{\sqrt{N}} \sum_{j=1}^N \ket{j}, 
\label{dd}
\end{equation}
where $\ket{j}$ is the site basis (one excitation on the $j$th molecule). 
All the other eigenvalues are zero, corresponding to a $(N-1)$-degenerate subspace spanned by all the states orthogonal to the lowest energy extended state.

Following~\cite{sokolov}, we can rewrite the Hamiltonian $H$ in Eq.~\eqref{Hfull} in the basis of these eigenstates, using the transformation matrix $U$, which has as columns the eigenstates of the matrix $Q$,
\begin{equation}
    H=U^T H_0 U - \frac{\gamma}{2} U^T Q U + \frac{\gamma}{2} I = \left(
\begin{array}{cc}
 -\dfrac{\gamma}{2}(N-1)+\zeta & \vec{h}^T \\
 \vec{h} & \tilde{H} \\
\end{array}
\right) \; .
\label{H-VQ}
\end{equation}
Let us note that the component $(1,1)$ of Eq.~\eqref{H-VQ} includes the term $\zeta= \sum_n \epsilon_n |\langle n|d\rangle|^2$, where $\epsilon_n$ and $\ket{n}$ are, respectively, eigenvalues and eigenvectors of $H_0$. 
Here, the matrix elements of the $(N-1) \times (N-1)$ submatrix $\tilde{H}$  on the basis of the excited states of the matrix  $ Q $ are
\begin{eqnarray}\label{Hmu0}
    \tilde{H}_{\mu \nu} &=& \braket{\mu|H_0 - \frac{\gamma}{2} Q + \frac{\gamma}{2} I|\nu} \nonumber \\
    &=& \braket{\mu|H_0|\nu} - \braket{\mu|\frac{\gamma}{2} Q|\nu} + \frac{\gamma}{2} \delta_{\mu \nu} \nonumber \; ;
\end{eqnarray}
here the second term vanishes since the eigenvalues of the degenerate excited states of $Q$ are $0$. 
If we rewrite the Hamiltonian $H_0$ in its eigenbasis $\ket{n}$, we get
\begin{eqnarray}\label{Hmu}
    \tilde{H}_{\mu \nu} &=& \bra{\mu} \left(\sum_n \epsilon_n \ket{n}\bra{n} \right) \ket{\nu} + \frac{\gamma}{2} \delta_{\mu \nu} \nonumber \\
    &=& \sum_n \epsilon_n \braket{\mu|n}   \braket{n|\nu} + \frac{\gamma}{2} \delta_{\mu \nu} \; .
\end{eqnarray}
The components of the vector $\vec{h}$, with dimension $N-1$, in the basis of the eigenstates of the matrix $Q$ are
\begin{equation}\label{hsmu}
    h_\mu = \sum_n \epsilon_n \braket{d|n}\braket{n|\mu} \;.
\end{equation}
The $\ket{\mu}$ eigenstates of $\tilde{H}$ are also eigenstates of $Q$ since they belong to the $(N-1)$-degenerate subspace of the $Q$ matrix. Thus, we diagonalize the submatrix $\tilde{H}$ and we call $\ket{\mu}$ its eigenstates with eigenvalues $\tilde\epsilon_\mu$, i.e.
\begin{equation}\label{Hmudiag}
    \tilde{H}_{\mu \nu} =  \tilde{\epsilon}_\mu \delta_{\mu \nu} .   
\end{equation}

If we multiply Eq.~\eqref{Hmu} by the components $\braket{m | \mu'}$,
where $\ket{\mu'}$ is an eigenstate of the submatrix $\tilde{H}$ and $\ket{m}$ is an eigenstate of $H_0$, and 
 we sum over all the states $\mu'$, we obtain
\begin{eqnarray}
    \sum_{\mu'} \tilde{H}_{\mu' \mu} \braket{m|\mu'} &=& \sum_{\mu'} \sum_n \epsilon_n \braket{\mu' | n}\braket{n|\mu}\braket{m | \mu'} + \frac{\gamma}{2} \sum_{\mu'} \delta_{\mu' \mu}  \braket{m | \mu'} 
    \nonumber \\
    &=& \sum_n \epsilon_n \braket{n|\mu} \bra{m}\left(\sum_{\mu'}  \ket{\mu'}\bra{\mu'} \right)\ket{n} + \frac{\gamma}{2} \braket{m|\mu}  \nonumber \\
    &=& \sum_n \epsilon_n \braket{n|\mu} \bra{m}\left( I-\ket{d}\bra{d} \right)\ket{n} + \frac{\gamma}{2} \braket{m|\mu} \nonumber \\
    &=& \sum_n \epsilon_n  \braket{n|\mu} \left( \delta_{m n}-\braket{m|d}\braket{d|n} \right) + \frac{\gamma}{2} \braket{m|\mu} \nonumber \\
     &=& (\epsilon_m + \frac{\gamma}{2} ) \braket{m|\mu} - \sum_n \epsilon_n \braket{d|n} \braket{n|\mu} \braket{m|d} \nonumber \\
     &=& (\epsilon_m + \frac{\gamma}{2} ) \braket{m|\mu} - h_\mu \braket{m | d} \; .
     \label{Hmu1}
\end{eqnarray}
On the other hand, from Eq.~\eqref{Hmudiag} we have
\begin{equation}
    \sum_{\mu'} \tilde{H}_{\mu' \mu} \braket{m | \mu'} = \sum_{\mu'} \tilde{\epsilon}_{\mu'} \delta_{\mu' \mu} \braket{m | \mu'} = \tilde{\epsilon}_\mu \braket{m|\mu}  \; .
    \label{Hmu2}
\end{equation}
By comparing Eqs.~\eqref{Hmu1} and~\eqref{Hmu2}, we obtain
\begin{equation}
    \tilde{\epsilon}_\mu \braket{m|\mu} = (\epsilon_m + \frac{\gamma}{2} ) \braket{m|\mu} - h_\mu \braket{m|d} \; ,
\end{equation}
i.e.,
\begin{equation}
    \ket{\mu} = \frac{h_\mu}{H_0 + \gamma/2 -\tilde{\epsilon}_\mu} \ket{d} \; .
    \label{eqmu}
\end{equation}
Equation~\eqref{eqmu} can be rewritten in the Anderson basis as follows
\begin{equation}
    \ket{\mu} = h_\mu \sum_n \frac{\braket{n|d}}{\epsilon_n + \gamma/2 -\tilde{\epsilon}_\mu} \ket{n} = \frac{h_\mu}{\sqrt{N}} \sum_n \frac{\sum_j \braket{n|j}}{\epsilon_n + \gamma/2 -\tilde{\epsilon}_\mu} \ket{n} \, ,
    \label{eqmu-f}
\end{equation}
where $\ket{j}$ is the site basis and the normalization coefficients $h_\mu$ are given by  
\begin{equation}
    h_\mu = \left( \sum_n \frac{\braket{d | n} \braket{n | d} }{(\epsilon_n + \gamma/2 -\tilde{\epsilon}_\mu)^2} \right)^{-1/2} \; .
    \label{hmu-norm}
\end{equation}

\begin{figure*}
    \centering
    \includegraphics[width=0.83\textwidth]{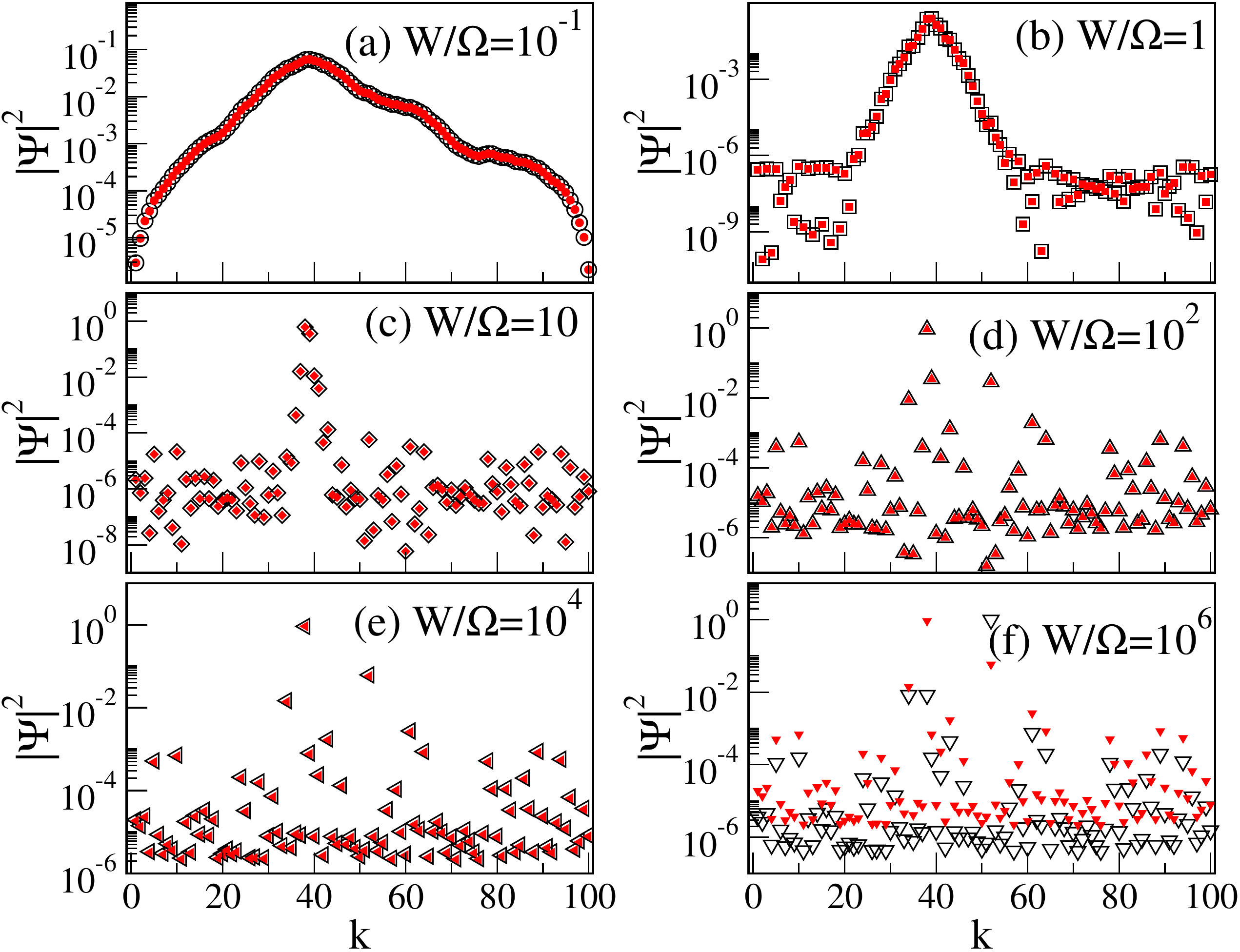}
    \vspace{-3mm}
    \caption{Eigenfunction probability $|\Psi|^2$ in the site basis $k$ for the first excited state (of a single disorder configuration) with disorder strengths $W/\Omega=\{10^{-1},1,10,10^2,10^4,10^6\}$. Here, $N=100$ and $\gamma=10^3\Omega$. The black symbols represent the exact eigenfunctions of the total Hamiltonian $H$ of Eq.~\eqref{HAcp} and the red symbols are the $\ket{\mu}$ states obtained from  Eq.~\eqref{eqmu-f}.}
    \label{Fig:10}
\end{figure*}

In the gapped regime and for sufficiently large disorder $W_2<W<W_{\rm GAP}$, where the eigenstates have a hybrid nature, we can assume that the Anderson eigenstates coincide with the site basis, see Fig.~\ref{Fig:Shape-H}(d). 
So, Eq.~\eqref{eqmu-f} with the normalization coefficients $h_\mu$ in Eq.~\eqref{hmu-norm} becomes
\begin{eqnarray}
    \ket{\mu} &\approx &  \displaystyle \left( \frac{1}{N} \sum_i \frac{1 }{(\epsilon_i + \gamma/2 -\tilde{\epsilon}_\mu)^2} \right)^{-1/2} \frac{1}{\sqrt{N}} \sum_i \frac{1}{\epsilon_i + \gamma/2 -\tilde{\epsilon}_\mu} \ket{i} \nonumber \\ 
  &  \approx & \displaystyle  \left( \sum_i \frac{1 }{\left( \frac{\epsilon_i-\tilde{\epsilon}_\mu }{W} + \frac{\gamma}{2W} \right)^2} \right)^{-1/2} \sum_i \frac{1}{\frac{\epsilon_i-\tilde{\epsilon}_\mu }{W} + \frac{\gamma}{2W}} \ket{i} \, .
  \label{hmu-gap2}
  \end{eqnarray}
  
Our perturbative expression allows to explain many features of the average shape of the eigenfunctions $\braket{|\Psi|^2}$ discussed above, mainly the existence of a disorder-independent plateau in the disorder range $W_2<W<W_{\rm GAP}$.
From Eq.~\eqref{hmu-gap2} we can see that both the normalization coefficients and the weights on the site basis are independent of disorder for $W \gg \gamma$ since both $\epsilon_i$ and $\tilde{\epsilon}_\mu$ are proportional to the disorder strength $W$ for large disorder. 
This is an important result since our perturbative approach is able to explain the independence of the plateau from disorder in the disorder regime $W_2<W<W_{\rm GAP}$. That is, since the disorder strength $W$ is uncorrelated from the site basis, the eigenfunction tail becomes a plateau that extends over the entire basis. Moreover, since the term $\frac{(\epsilon_i-\tilde{\epsilon}_\mu )}{W}$ decreases as $1/N$, this also explains the dependence of the probability of the extended tails on the system size for $W \gg \gamma$.

Finally, to validate the perturbative derivation of the $\ket{\mu}$ states above, in Fig.~\ref{Fig:10} we present the eigenfunction probability $|\Psi|^2$, in the site basis $k$, for the first excited state at several disorder strengths $W$. In each panel we show the exact eigenfunction of the total Hamiltonian $H$ of Eq.~\eqref{HAcp} (black symbols) and the $\ket{\mu}$ state obtained from Eq.~\eqref{eqmu-f} (red symbols). In all panels from (a) to (e) we see an excellent correspondence between the exact and the $\ket{\mu}$ state. Note that since  $W_{\rm GAP}\approx 230258$ for the parameter chosen in Fig.~\ref{Fig:10}, we cannot expect agreement in panel (f).

Using the perturbative expressions of the eigenstates obtained in Eqs.~\eqref{dd} and \eqref{eqmu-f} we can compute both the current through Eq.~(\ref{eq4}) and the variance. The results are shown in Fig.~2(a,b) of the Main Text, see orange squares.

\section{Master Equation {\it vs.} Schr\"odinger equation approach to compute the current}

\begin{figure*}[t]
    \centering
    \includegraphics[width=0.84\textwidth]{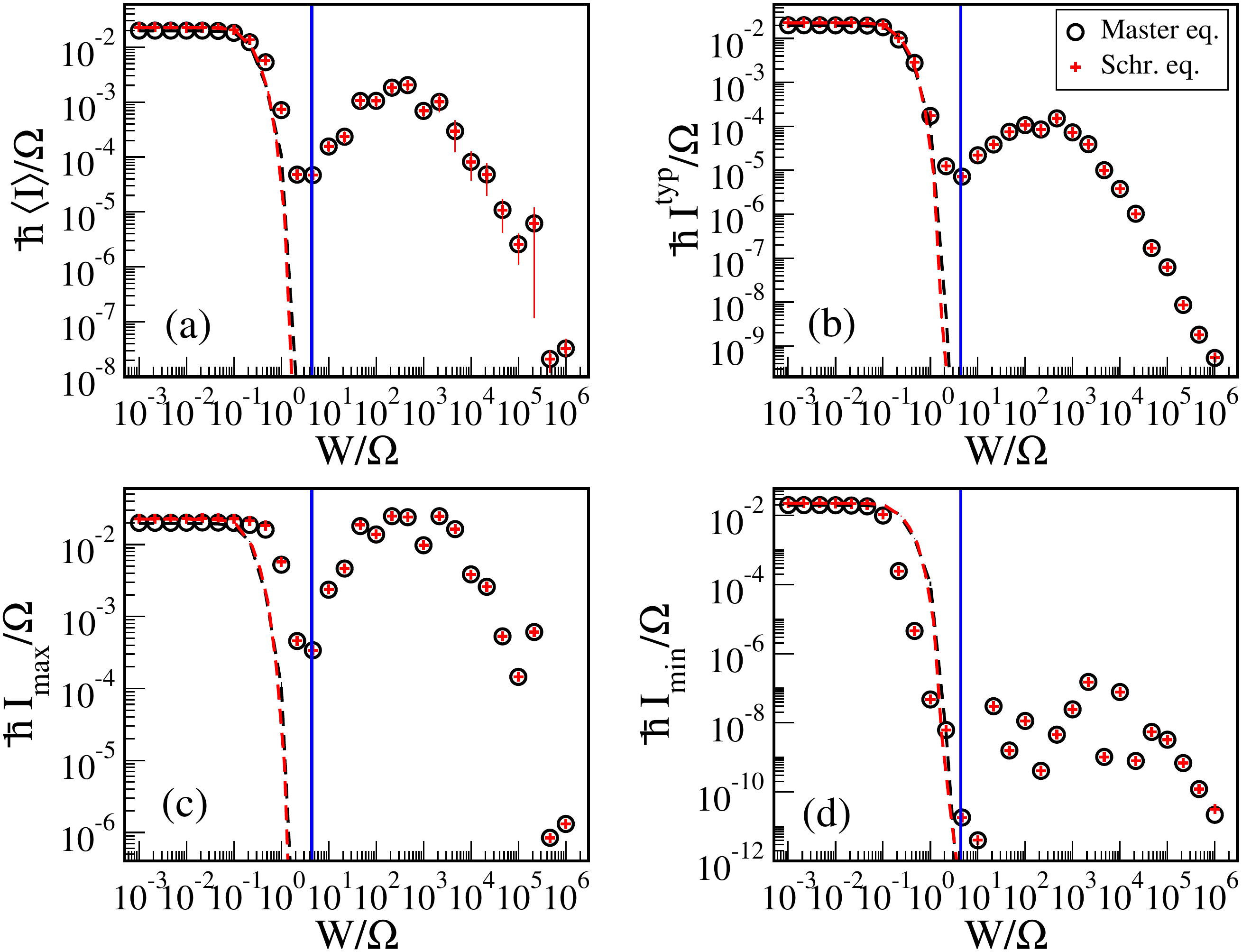}
    \vspace{-3mm}
    \caption{(a) Average $\braket{I}$, 
    (b) typical $I^{\rm typ}$, 
    (c) maximal $I_{max}$ and (d) minimal $I_{min}$ currents, multiplied by $\hbar/\Omega$, as a function of the normalized static disorder $W/\Omega$ for the long-range Hamiltonian, Eq.~\eqref{HAcp}. The  stationary current is computed with the master equation approach (open circles), see Eqs.~(4,5) in the Main Text, and with the non-Hermitian Schr\"odinger equation equation approach (red crosses), see Eqs.~(6-9) in the Main Text. Here, $N=40$,
    $\gamma_d=\gamma_p=\Omega$, $\gamma=10 \,\Omega$. The averages are taken over $N_r=100$ disorder configurations.
    The dashed curves indicate the case $\gamma=0$.
    Vertical blue lines indicate the critical disorder $W_1$ given by Eq.~\eqref{Wcr_1}.}
    \label{fig:4a}
\end{figure*}

\begin{figure*}[!hb]
    \centering
    \includegraphics[width=0.84\textwidth]{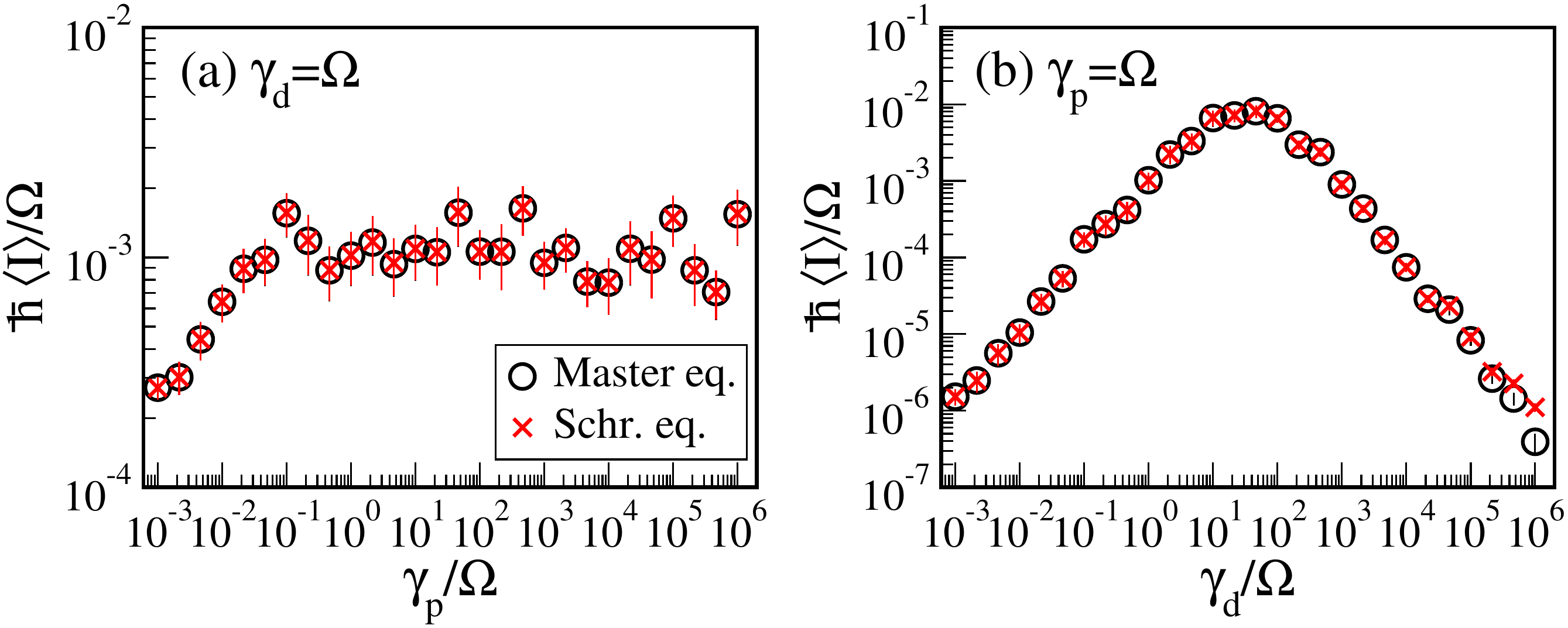}
    \vspace{-4mm}
    \caption{(a) Normalized average current $\hbar \braket{I}/\Omega$ {\it vs.} the normalized coupling strength $\gamma_p/\Omega$ for $\gamma_d=\Omega$ and (b) normalized average current $\hbar \braket{I}/\Omega$ {\it vs.} the normalized coupling strength $\gamma_d/\Omega$ for $\gamma_p=\Omega$. The  stationary current is computed with the master equation approach (open circles), see Eqs.~(4,5) in the Main Text, and with the non-Hermitian Schr\"odinger equation equation approach (red crosses), see Eqs.~(6-9) in the Main Text. Here, $W=100\,\Omega$, while the other parameters are the same as in Fig.~\ref{fig:4a}.} 
    \label{fig:4b}
\end{figure*}

In the Main Text we report the transport properties of two models: the 1D Anderson model subject to long-range hopping and a disordered molecular wire placed in an optical cavity. For both systems we report the stationary current as the main figure of merit to characterize transport. However, the standard master equation approach (see Eqs.~(4,5) in the Main Text) is numerically very expensive to compute the stationary current through long wires. For this reason, we use a different definition of current that is based on a non-Hermitian Schr\"odinger equation (see Eqs.~(6-9) in the Main Text) which is computationally less expensive. Here we add some details about the non-Hermitian Schr\"odinger equation approach and we analytically  prove  the identity between the two definitions of the current.

\subsection{Non-Hermitian Schr\"odinger equation approach}

The non-Hermitian Schr\"odinger equation approach described in the Main Text (see Eqs.(6-9)) is based on the calculation of the average escape time from the chain, when the excitation is initialized on the site $\ket{1}$ and in presence of a drain at site $\ket{N}$. Such drain is described by the effective Hamiltonian (see Eq.~(7) in the Main Text)
\begin{equation}
    (H_{\rm eff})_{k,l}= (H)_{k,l}-i\,\frac{\gamma_d}{2}\, \delta_{k,N}\delta_{l,N},
    \label{Hefftau}
\end{equation}
where $H$ is the Hamiltonian from Eq.~\eqref{HAcp} and $\delta_{k,N}$ is the Kronecker delta.

The average transfer time $\tau$ is defined as (see Eq.~(6) in the Main Text)
\begin{equation}
    \tau=\frac{\gamma_d}{\hbar} \int_0^\infty dt \,t \, |\Psi_N(t)|^2 ,
    \label{eq1}
\end{equation}
with $\Psi_N(t)=\bra{N}e^{-iH_{\rm eff} t/\hbar}\ket{1}$ being the probability amplitude on the drain site of a time-evolved wave function at time $t$, under the effective Hamiltonian $H_{\rm eff}$ of Eq.~\eqref{Hefftau}. The integral in Eq.~\eqref{eq1} can be evaluated analytically by expanding $e^{-iH_{\rm eff} t/\hbar}$ on the eigenbasis of $H_{\rm eff}$ which, being non-Hermitian, has right and left eigenvectors,
\begin{equation}
\label{biorth}
    H_{\rm eff} \ket{r_k} = \varepsilon_k \ket{r_k} \qquad {\rm and} \qquad \bra{\tilde{r}_k} H_{\rm eff} = \bra{\tilde{r}_k} \varepsilon_k~.
\end{equation}

Performing the integral one gets for the average transfer $\tau$ the following expression, 
%
%
\begin{equation}
    \tau = \hbar\gamma_d \sum_{k,k'} \frac{\braket{N|r_k} \braket{{\tilde r}_k|1} \braket{N|r_{k'}}^* \braket{{\tilde r}_{k'}|1}^*}{-(\varepsilon_k-\varepsilon_{k'}^*)^2}~,
    \label{eq4}
\end{equation}
which depends only on the eigenvalues and eigenvectors of $H_{\rm eff}$ and it is used in Eqs.~(8,9) of the Main Text to compute the steady-state current.

\subsection{Exact mapping between master equation and Schr\"odinger equation approaches}

Here we consider the master equation, Eq.~(4) of the Main Text, and we prove that the steady-state current, defined in Eq.~(5) of the Main Text, is identical to the Schr\"odinger equation result, see Eq.~(9) in the Main Text. The master equation, Eq.~(4) of the Main Text, is written explicitly as
\begin{equation}
\label{mefull}
    \frac{d\rho}{dt} = -\frac{i}{\hbar} \left( H \rho - \rho H \right) - \frac{\gamma_d}{2\hbar} \left( \ket{N}\bra{N} \rho + \rho \ket{N}\bra{N} \right) + \frac{\gamma_d}{\hbar} \rho_{NN} \ket{0}\bra{0} - \frac{\gamma_p}{2\hbar} \left( \ket{0}\bra{0}\rho + \rho\ket{0}\bra{0} \right) + \frac{\gamma_p}{\hbar} \rho_{00}\ket{1}\bra{1} ,
\end{equation}
where $H$ is a generic hermitian Hamiltonian acting on the single-excitation subspace, $\gamma_d/\hbar$ is the drain rate from the site $\ket{N}$, $\rho_{NN}=\braket{N|\rho|N}$ is the population of the site $\ket{N}$, $\ket{0}$ is the vacuum state, $\gamma_p/\hbar$ is the pumping rate on the site $\ket{1}$ and $\rho_{00}=\braket{0|\rho|0}$ is the population of the vacuum state. First, we note that Eq.~\eqref{mefull} can be written in terms of the effective Hamiltonian, Eq.~\eqref{Hefftau}, and it reads
\begin{equation}
\label{meHeff}
    \frac{d\rho}{dt} = -\frac{i}{\hbar} \left( H_{\rm eff} \rho - \rho H_{\rm eff}^\dag \right) + \frac{\gamma_d}{\hbar} \rho_{NN} \ket{0}\bra{0} - \frac{\gamma_p}{2\hbar} \left( \ket{0}\bra{0}\rho + \rho\ket{0}\bra{0} \right) + \frac{\gamma_p}{\hbar} \rho_{00}\ket{1}\bra{1}~.
\end{equation}

We want to compute the steady-state current, which is defined in Eq.~(5) of the Main Text as
\begin{equation}
\label{Iss}
    I = \frac{\gamma_d}{\hbar} \braket{N|\rho^{(ss)}|N},
\end{equation}
where $\rho^{(ss)}$ is the steady-state density matrix, that we obtain by setting $\frac{d\rho}{dt}=0$ in Eq.~\eqref{meHeff}. First, we set the derivative of the vacuum state population to zero, {\it i.e.}
\begin{equation}
\label{drho00}
    \frac{d}{dt}\braket{0|\rho^{(ss)}|0} = \frac{\gamma_d}{\hbar} \rho_{NN}^{(ss)} - \frac{\gamma_p}{\hbar} \rho_{00}^{(ss)} = 0~,
\end{equation}
where we used the fact that $H_{\rm eff}$ acts only on the single-excitation subspace, {\it i.e.} $H_{\rm eff}\ket{0}=0$.
From Eq.~\eqref{drho00} we have $\gamma_d\rho_{NN}^{(ss)}=\gamma_p\rho_{00}^{(ss)}$, so that the steady-state current, Eq.~\eqref{Iss} can be expressed as
\begin{equation}
\label{Iss00}
    I = \frac{\gamma_d}{\hbar} \rho^{(ss)}_{NN} = \frac{\gamma_p}{\hbar} \rho^{(ss)}_{00}~.
\end{equation}
Now we proceed to compute $\rho^{(ss)}_{00}$, and we use the fact that the total trace of the density matrix must be unity, {\it i.e.}
\begin{equation}
\label{trss}
    {\rm Tr}[\rho^{(ss)}] = \rho_{00}^{(ss)} + \sum_j \braket{j|\rho^{(ss)}|j} = 1~,
\end{equation}
where the states $\ket{j}$ form a generic orthonormal basis on the single-excitation subspace. Now, we recall that the eigenstates of the effective Hamiltonian $H_{\rm eff}$, see Eq.~\eqref{biorth}, form a biorthogonal basis, {\it i.e.} $\braket{\tilde{r}_k|r_{k'}}=\delta_{k,k'}$. This allows to decompose the identity (on the single-excitation subspace) as
\begin{equation}
    {\rm Id_{s.e.s.}} = \sum_k \ket{r_k}\bra{\tilde{r}_k} = \sum_k \ket{\tilde{r}_k}\bra{r_k}~. 
\end{equation}
Using the above decompositions, we can express the sum over $j$ in Eq.~\eqref{trss} as
\begin{equation}
\label{sumjj}
    \sum_j \braket{j|\rho^{(ss)}|j} = \sum_j\sum_k\sum_{k'} \braket{j|r_k} \braket{\tilde{r}_k|\rho^{(ss)}|\tilde{r}_{k'}} \braket{r_{k'}|j} = \sum_k\sum_{k'} \braket{r_{k'}|r_k} \braket{\tilde{r}_k|\rho^{(ss)}|\tilde{r}_{k'}}~.
\end{equation}
Here above we also used the fact that $\ket{j}$ is an orthonormal basis on the single-excitation subspace, so that it is possible to decompose the scalar product $\braket{r_k|r_{k'}}=\sum_j\braket{r_{k'}|j}\braket{j|r_k}$. Note that the eigenstates $\ket{r_k}$ are not orthonormal, so that $\braket{r_k|r_{k'}}\neq 0$ for $k\neq k'$. Specifically, we can compute $\braket{r_k|r_{k'}}$ as follows. From the non-Hermitian Hamiltonian, see Eq.~\eqref{Hefftau}, we have the identity
\begin{equation}
    H_{\rm eff} - H_{\rm eff}^\dag = -i \gamma_d \ket{N}\bra{N}~.
\end{equation}
If we take the expectation value of both sides of the above equation between $\braket{r_{k'}|\dots|r_k}$, using Eq.~\eqref{Hefftau}, we obtain
\begin{equation}
    \left( \varepsilon_k - \varepsilon_{k'}^* \right) \braket{r_{k'}|r_k} = -i\gamma_d \braket{r_{k'}|N} \braket{N|r_k}
\end{equation}
from which we have
\begin{equation}
\label{scalkk}
    \braket{r_{k'}|r_k} = \frac{\gamma_d \braket{r_{k'}|N} \braket{N|r_k}}{i\left( \varepsilon_k - \varepsilon_{k'}^* \right)}~.
\end{equation}
Now, to evaluate Eq.~\eqref{sumjj}, we proceed to compute $\braket{\tilde{r}_k|\rho^{(ss)}|\tilde{r}_{k'}}$ by setting $\frac{d\rho}{dt}=0$. Specifically, from Eq.~\eqref{meHeff}, using Eq.~\eqref{Hefftau}, we obtain
\begin{equation}
    \frac{d}{dt}\braket{\tilde{r}_k|\rho^{(ss)}|\tilde{r}_{k'}} = -\frac{i}{\hbar} \left( \varepsilon_k - \varepsilon_{k'}^* \right) \braket{\tilde{r}_k|\rho^{(ss)}|\tilde{r}_{k'}} + \frac{\gamma_p}{\hbar} \rho_{00}^{(ss)} \braket{\tilde{r}_k|1}\braket{1|\tilde{r}_{k'}} = 0~,
\end{equation}
from which we have
\begin{equation}
\label{rhokk}
    \braket{\tilde{r}_k|\rho^{(ss)}|\tilde{r}_{k'}} = \gamma_p \rho_{00}^{(ss)} \frac{\braket{\tilde{r}_k|1}\braket{1|\tilde{r}_{k'}}}{i\left( \varepsilon_k - \varepsilon_{k'}^* \right)}~.
\end{equation}

Now, we substitute Eq.~\eqref{scalkk} and Eq.~\eqref{rhokk} into Eq.~\eqref{sumjj} and obtain
\begin{equation}
    \sum_j \braket{j|\rho^{(ss)}|j} = \frac{\gamma_p\rho_{00}^{(ss)}}{\hbar} \left[ \hbar \gamma_d \sum_{k,k'} \frac{\braket{r_{k'}|N} \braket{N|r_k}\braket{\tilde{r}_k|1}\braket{1|\tilde{r}_{k'}}}{-\left( \varepsilon_k - \varepsilon_{k'}^* \right)^2} \right]~.
\end{equation}
Note that the term inside square brackets is equal to the average transfer time $\tau$, see Eq.~\eqref{eq4}, {\it i.e.}
\begin{equation}
\label{sumjjtau}
    \sum_j \braket{j|\rho^{(ss)}|j} = \frac{\gamma_p\tau\rho_{00}^{(ss)}}{\hbar}~.
\end{equation}
Therefore, by substituting Eq.~\eqref{sumjjtau} into Eq.~\eqref{trss} we obtain
\begin{equation}
    \rho_{00}^{(ss)} + \frac{\gamma_p\tau}{\hbar}\rho_{00}^{(ss)} = 1~,
\end{equation}
from which we obtain the steady-state value of the population of the vacuum state,
\begin{equation}
\label{rho00ss}
    \rho_{00}^{(ss)} = \frac{1}{1 + \frac{\gamma_p\tau}{\hbar}}~.
\end{equation}

Finally, we substitute Eq.~\eqref{rho00ss} into Eq.~\eqref{Iss00} and we have
\begin{equation}
    I = \frac{\gamma_p}{\gamma_p\tau + \hbar}~,
\end{equation}
which is exactly the value of the steady-state current that we obtained with the non-Hermitian Schr\"odinger equation approach, see Eq.~(9) in the Main Text. Note that in our calculations we did not specify the nature of the hermitian Hamiltonian $H$, so our results work for a general open quantum system in the single-excitation approximation, with incoherent pumping of excitation on one state of the system (state $\ket{1}$) and incoherent draining of excitation from another state (state $\ket{N}$).

In the following, we compare  analytical  with numerical results, and we show that the master equation approach gives a steady-state current  identical to that obtained via the Schr\"odinger equation.

In Fig.~\ref{fig:4a} we plot the current  multiplied by $\hbar/\Omega$ as a function of the normalized static disorder $W/\Omega$ for the long-range Hamiltonian, on a chain of $N=40$ sites, from both approaches: the master equation approach (black symbols) and the non-Hermitian Schr\"odinger equation approach (red symbols). We are limited to consider a short chain of $N=40$ sites because the master equation approach is numerically very expensive. Moreover, we set an unusually large value of the long-range coupling ($\gamma=10\,\Omega$), to ensure that the disorder threshold $W_{\rm GAP}$ (see Eq.~\eqref{WGAP}) is larger than the disorder threshold $W_1$ (see Eq.~\eqref{Wcr_1}) even for such a small system size.
Specifically, in the panels of Fig.~\ref{fig:4a} we report: the average $\braket{I}$, Fig.~\ref{fig:4a}(a); the typical $I^{\rm typ}$, Fig.~\ref{fig:4a}(b); the maximal $I_{max}$, Fig.~\ref{fig:4a}(c) and the minimal  $I_{min}$ currents, Fig.~\ref{fig:4a}(d); all of them are multiplied by $\hbar/\Omega$. In all cases, we observe a perfect match between the two approaches, thus validating the use of the non-Hermitian Schr\"odinger equation approach in the Main Text. 
Note that the large error bar for the average current, Fig.~\ref{fig:4a}(a), present for $W> 10^5 \Omega$ can be explained by an anomalously large value of the current in one of the 100 disorder realizations used to produce the figure, as we verified. Probably, a larger number of realizations
would fix this problem, but we are not interested in that since our main results are about the typical current, which is self-averaging and does not present this problem, see Fig.~\ref{fig:4a}(b).

Moreover, in Fig.~\ref{fig:4b} we report the normalized average current $\hbar \braket{I}/\Omega$ from the master equation approach (black symbols) and the non-Hermitian Schr\"odinger equation approach (red symbols) as a function of the normalized pumping rate $\gamma_p/\Omega$ (for fixed $\gamma_d$), see Fig.~\ref{fig:4b}(a), and of the normalized draining rate $\gamma_d/\Omega$ (for fixed $\gamma_p$), see Fig.~\ref{fig:4b}(b), for a fixed value of disorder ($W=100\Omega$).
Similarly to Fig.~\ref{fig:4a}, also here we observe a perfect correspondence between the two approaches.

\section{Current and typical current}

 \begin{figure*}[!htb]
    \centering 
    \includegraphics[width=0.72\textwidth]{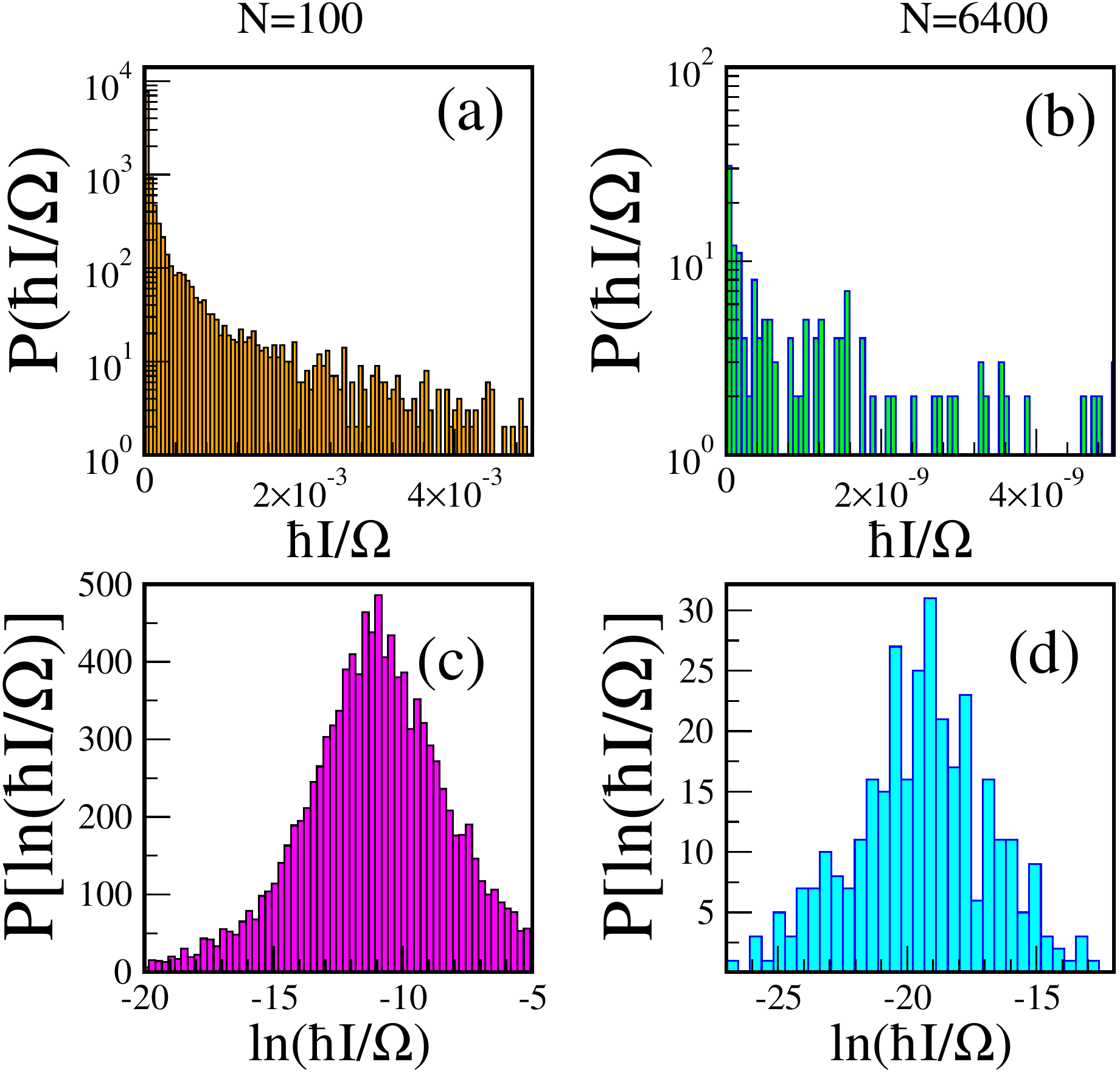}
    \vspace{-3mm}
    \caption{Probability distributions of (a,b) the normalized current $\hbar I/\Omega$ and (c,d) the logarithm of the normalized current $\ln (\hbar I/\Omega)$, for two wire sizes (a,c) $N=100$ and (b,d) $N=6400$. We considered the 1D Anderson model with long-range hopping, see Eq.~\eqref{HAcp}, with $\gamma=\Omega$ and $W=100~\Omega$. The number of random disorder configurations is (a,c) $N_r=10^4$ and (b,d) $N_r=350$.}
    \label{Fig:20}
    \end{figure*}

In Fig.~\ref{Fig:20} the probability distribution functions (PDF)  of the stationary current $\hbar I/\Omega$ and of the variable $J= \ln (\hbar I/\Omega)$ are shown for a case in the disorder-independent transport (DIT)  regime. As one can see, while the distribution of the current is strongly peaked at the origin and develops a slow-decaying tail, the distribution of $J$ has a bell shape.

\begin{figure*}[!t]
    \centering 
    \includegraphics[width=0.83\textwidth]{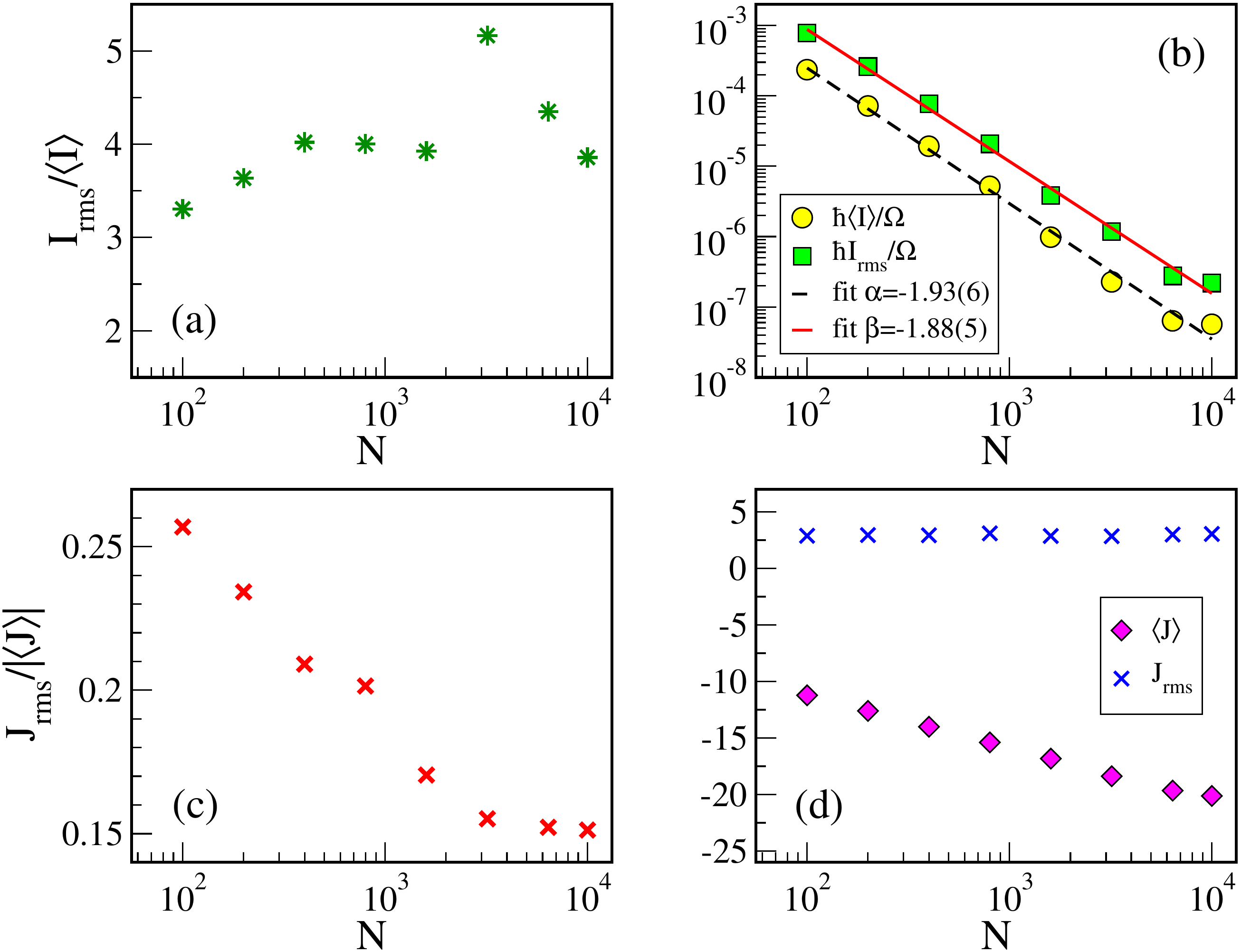}
    \vspace{-4mm}
    \caption{(a) The ratio $I_{rms}/\langle I\rangle$, (b) $\langle I \rangle$ and $I_{rms}$ in units of the hopping rate $(\Omega/\hbar)$, (c) the ratio $J_{rms}/|\langle J\rangle |$, and (d) $\langle J \rangle$ and $J_{rms}$ as a function of the wire size $N$.
    We considered the 1D Anderson model with long-range hopping, see Eq.~\eqref{HAcp}, with $\gamma=\Omega$ and $W=100~\Omega$. The full-red and back-dashed lines in panel (b) are power-law fittings to the data with exponents $\alpha, \beta$ as given in the legend.
    The number of disorder configurations $N_r$ is such that $N \times N_r = 10^6$.}
    \label{Fig:21}
    \end{figure*}

Therefore it is important to check which of the two quantities has the self-averaging property, namely, a  ratio between the standard deviation $I_{rms}$ and the mean $\langle I\rangle$  decreasing with $N$ for large $N$ values. Results for the variables $I$ and $J$ are shown in Figs.~\ref{Fig:21}(a,c). As one can see, while in the first case 
$I_{rms}/\langle I\rangle$ grows with $N$, $J_{rms}/|\langle J\rangle|$ decreases with $N$. For this reason we decided to consider the variable $J$, and from that the typical current $\hbar I^{\rm typ}/\Omega=\exp(\langle J\rangle)$, in most of our numerical calculations in the Main Text.


   \begin{figure*}
    \centering 
    \includegraphics[width=0.38\columnwidth]{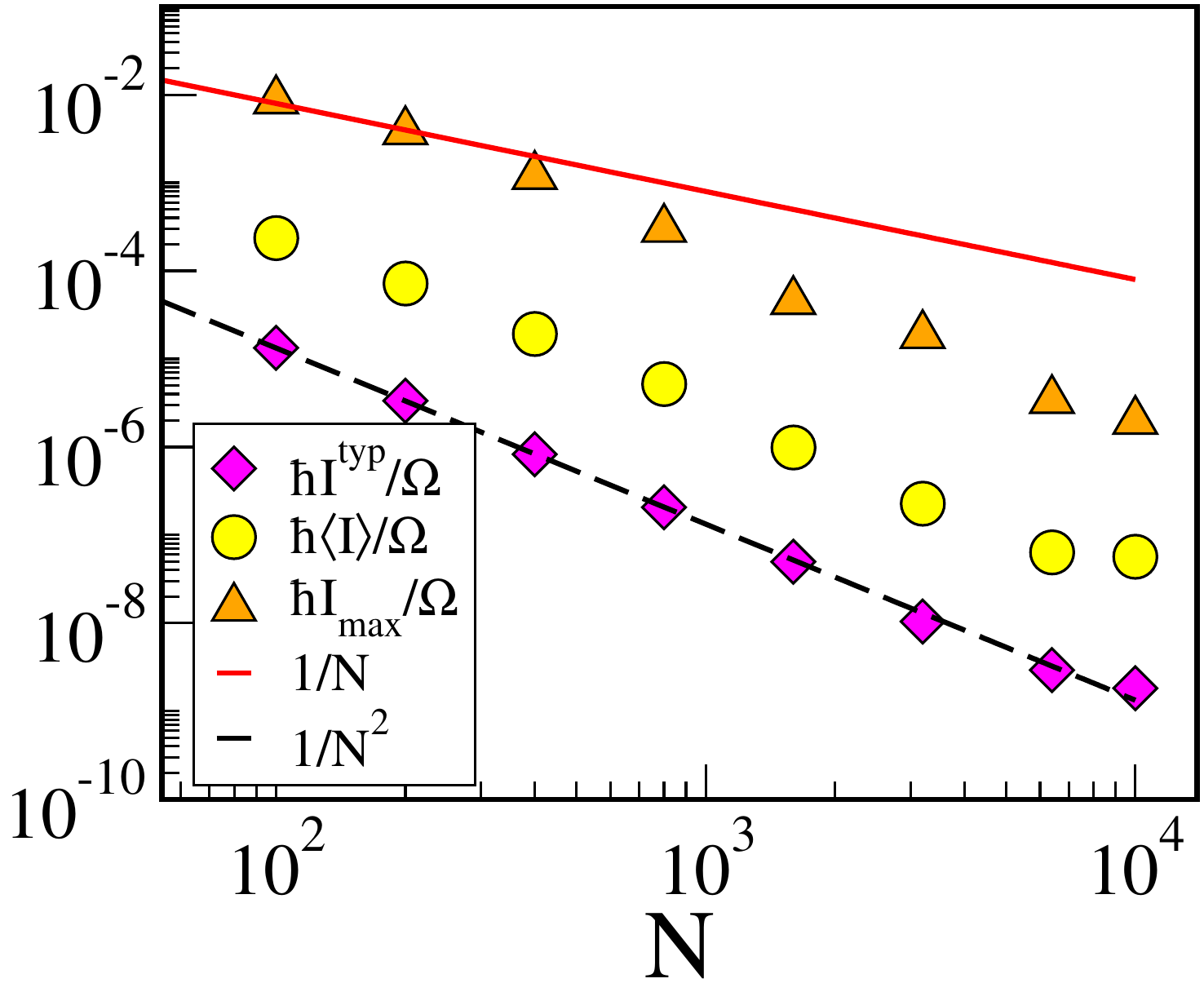}
    \vspace{-4mm}
    \caption{Behavior of average $\braket{I}$, typical $I^{\rm typ}$ and maximal $I_{max}$ currents (in units of the hopping rate $\Omega/\hbar$) in the disorder-independent regime as a function of the system size $N$ for an all-to-all coupling using the same parameters as Fig.~\ref{Fig:21}. The full-red and dashed-black lines are power-laws of $N$ as indicated in the legend.}
    \label{Fig4x}
    \end{figure*}

For completeness in Fig.~\ref{Fig:21}(b) we show the average current $\braket{I}$ and its standard deviation $I_{rms}$ as a function of the wire size $N$. Figure~\ref{Fig:21}(d) is equivalent to Fig.~\ref{Fig:21}(b) but for the variable $J$.
Moreover, in Fig.~\ref{Fig4x} we  compare the behavior of 
the average $\braket{I}$, typical $I^{\rm typ}$ and maximal $I_{max}$ currents, in the disordered-independent regime, as a function of the system size $N$. We observe that all currents reported in Fig.~\ref{Fig4x} are proportional to $1/N^2$, see the dashed line; this size dependence was already reported in Fig.~\ref{Fig:21}(b) for the average current $\braket{I}$.

\section{Scaling of the transport regimes with the long-range coupling strength}

\begin{figure}[!ht]
    \centering 
    \includegraphics[width=0.85\textwidth]{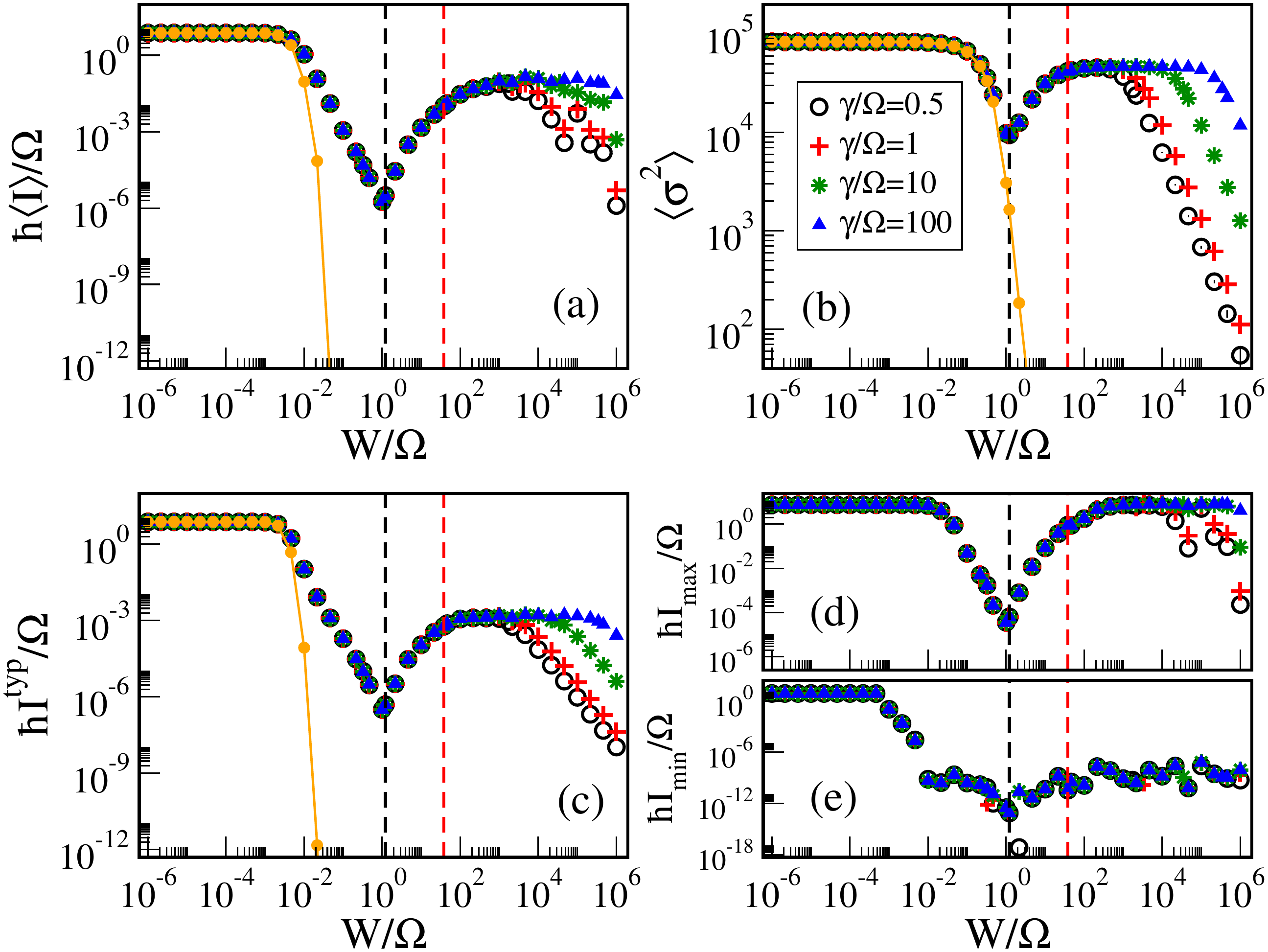}
    \vspace{-2mm}
    \caption{(a) Average $\braket{I}$, (c) typical $I^{\rm typ}$, (d) maximal $I_{max}$ and (e) minimal $I_{min}$ currents in units of the hopping rate $\Omega/\hbar$, and (b) average variance $\braket{\sigma^2}$ as a function of the static disorder $W$ for different coupling strengths $\gamma$ as indicated in the legend. Here, $N=10^3$ and $\gamma_p=\gamma_d=\Omega$. We used $N_r= 10^3$ disorder configurations.  
    The orange curves show the case $\gamma=0$ while the vertical dashed lines indicate the critical disorder $W_1$ (black) and $W_2$ (red).}
    \label{Fig:14}
    \end{figure}

In Fig.~\ref{Fig:14} we present the average $\braket{I}$, typical $I^{\rm typ}$, maximal $I_{max}$ and minimal $I_{min}$ currents, multiplied by $\hbar/\Omega$, as a function of the normalized static disorder $W/\Omega$ for disordered wires with long-range hopping with different coupling strengths $\gamma$, as indicated in panel (b). 
For comparison purposes in Fig.~\ref{Fig:14}(b) we also report the average variance $\braket{\sigma
^2}$ of the excited eigenstates. As a reference, in panels (a-c) the case of $\gamma=0$ is also shown, see the orange curves. In all panels the values of $W_1$ and $W_2$ are indicated with vertical dashed lines.

From Fig.~\ref{Fig:14} there are some points that deserve to be highlighted:
$(i)$ Since we use a fixed wire size, $N=10^3$, all curves fall one on the top of the other for $W<W_{\rm GAP}$; recall that neither $W_1$ nor $W_2$ depend on the coupling strength $\gamma$, see Eqs.~\eqref{Wcr_1} and~\eqref{W2_2}, respectively.
$(ii)$ Since $W_{\rm GAP}\propto \gamma$, see Eq.~\eqref{WGAP}, the larger the value of the coupling strength $\gamma$ the wider the disorder independent regime $W_2<W<W_{\rm GAP}$. It is interesting to note that transport properties do not depend on the value of the long range coupling $\gamma$ in the gapped regime.

\section{Mapping between a molecular chain in an optical cavity and a system with long-range hopping}

Here we consider a molecular chain in an optical cavity, with an optical mode  at resonance with the molecule excitation energy. We show that the common coupling to the cavity mode effectively induces a long-range hopping between the  molecules. 
This mapping, see Fig.~3(a) in the Main Text, is very accurate even in presence of disorder until $W \approx W_{\rm GAP}$. 

The molecular chain in the cavity is described by the Hamiltonian in Eq.~(12) of the Main Text. 
First we proceed to prove that in absence of disorder there is an energy gap $\Delta$ between the polaritonic ground state and the lowest energy excited state. 
Indeed, in absence of disorder ($W=0$), only the fully symmetric state $\ket{d}$, see Eq.~\eqref{dd}, with energy $-2 \Omega$ in the molecular chain couples with strength $\sqrt{N}g$ with the cavity mode which is at energy zero (resonant with the molecule excitation energy). Thus, we can compute the polaritonic energies and the energy gap $\Delta$ by solving the $2 \times 2$ coupling matrix for $\ket{d}$ and the cavity mode $\ket{c}$,
\[
\begin{pmatrix}
    \bra{d} H_{cav} \ket{d}     & \bra{d} H_{cav} \ket{c} \\
    \bra{c} H_{cav} \ket{d}   & \bra{c} H_{cav} \ket{c}
\end{pmatrix}
=
\begin{pmatrix}
    -2\Omega     & \sqrt{N}g \\
    \sqrt{N}g   & 0
\end{pmatrix}~,
\]
with $H_{cav}$ from Eq.~(12) in the Main Text. Considering that for large $N$ also the first excited state has energy $-2 \Omega$, we have
\begin{equation}
\Delta=\sqrt{Ng^2+\Omega^2}-\Omega \approx \sqrt{N}g \quad \mbox{for} \quad \sqrt{N}g \gg \Omega \, .
\label{gapC}
\end{equation}

On the other hand, a molecular chain in presence of long-range hopping of coupling strength $\gamma$ has an energy gap $\Delta$, in absence of disorder, equal to $N\gamma/2$. By imposing $\gamma_{\rm eff}= 2g /\sqrt{N}$ we determine the effective long-range coupling which would produce the same energy gap $\Delta$ in absence of disorder and for large $N$. With this choice of $\gamma_{\rm eff}$, if we exclude the polaritonic states in the cavity model and the ground state in the long range model, all other eigenstates and eigenvalues between the two models are identical for $W=0$.   This does not prove that they will be equivalent when disorder is added. In order to discuss this point, in the following, we consider the role of disorder using perturbation theory.

By considering the Anderson model,
 $$H_0 = \sum_n \epsilon_n \ket{n}\bra{n}$$
(see also Eq.~(2) in the Main Text)
as a perturbation in the regime $\sqrt{N}g \gg (\Omega, W$) or $\gamma_{\rm eff} \gg (\Omega, W)$ and following the same approach developed in previous Sections,
we can apply a perturbative approach to both the long-range and the cavity model.

Let us start to consider the long-range Hamiltonian. 
We can write the  Hamiltonian $H$ in Eq.~(1) of the Main Text in the basis which diagonalizes the long-range interaction matrix $V=-\gamma/2 \sum_{i \ne j} \ket{i}\bra{j}$. 
This basis is formed by the fully symmetric state $\ket{d}$, with eigenvalue $-\gamma (N-1)/2$, and by the $N-1$ degenerate states $\ket{\mu}$, orthogonal to $\ket{d}$, with eigenvalues $\gamma/2$. In this basis  we have
\begin{equation}
    H=  \left(
\begin{array}{cc}
 -\dfrac{\gamma}{2}(N-1) +\zeta & \vec{h}^T \\
 \vec{h} & \tilde{H} \\
\end{array}
\right) \; ,
\label{HL}
\end{equation}
where $\vec{h}$ is the interaction vector between the excited states and the ground state~\eqref{hmu-norm} of the long-range interaction matrix $V$ with components
\begin{equation}
h_{\mu}= \sum_n \epsilon_n \langle d|n\rangle \langle n|\mu \rangle \, .
\label{hu2}
\end{equation}
Moreover,
\begin{equation}
\zeta= \sum_n \epsilon_n |\langle n|d\rangle|^2
\label{zeta}    
\end{equation}
and $\tilde{H}$ represents the matrix elements of the $N-1$ excited states written in the  degenerate basis of the long-range hopping interaction matrix.
The matrix elements of $\tilde{H}$ with respect to the $N-1$ degenerate eigenstates of $V$, $\ket{\mu},\ket{\nu}$, can be written as
\begin{equation}
\langle \nu|\tilde{H}| \mu \rangle= \sum_n  \epsilon_n    \langle \nu|n\rangle \langle n|\mu \rangle +\frac{\gamma}{2}\delta_{\nu\mu}~.
\label{Htilde}    
\end{equation}

For the molecular chain in the cavity we can use as a basis the eigenstates of the interaction matrix between the $N$ uncoupled molecules and the cavity mode. 
In this interaction matrix all the molecules are coupled with strength $g$ with the cavity mode which acts as an additional external site, see Eq.~(12) in the Main Text. This form of the coupling implies that only the state $\ket{d}$ in the molecular chain couples with the cavity mode, forming two polaritonic states $\ket{p_\pm}$  with energies $\pm \sqrt{N} g$. On the other hand all the other $N-1$ degenerate eigenstates $\ket{\mu}$ with energy zero are decoupled from the cavity mode. Note that the $\ket{\mu}$ states defined here are identical to the $\ket{\mu}$ states defined above for the long-range model.   Therefore,
in the basis $\left\{ \ket{p_\pm},\ket{\mu}\right\}$ we can write the Hamiltonian of Eq.~(12) in the Main Text as: 
\begin{equation}
    H=  \left(
\begin{array}{ccc}
 -\sqrt{N}g +\zeta/2 &\zeta/2 & \vec{h}^T/2 \\
 \zeta/2 & \sqrt{N}g+\zeta/2 & \vec{h}^T/2 \\
  \vec{h}/2 & \vec{h}/2 & \tilde{H} \\
\end{array}
\right) \; ,
\label{HC}
\end{equation}
where the matrix elements $\vec{h}$ and $\tilde{H}$ are given above, see Eqs.~(\ref{hu2},\ref{zeta},\ref{Htilde}).

\begin{figure*}[!t]
    \centering 
    \includegraphics[width=0.65\textwidth]{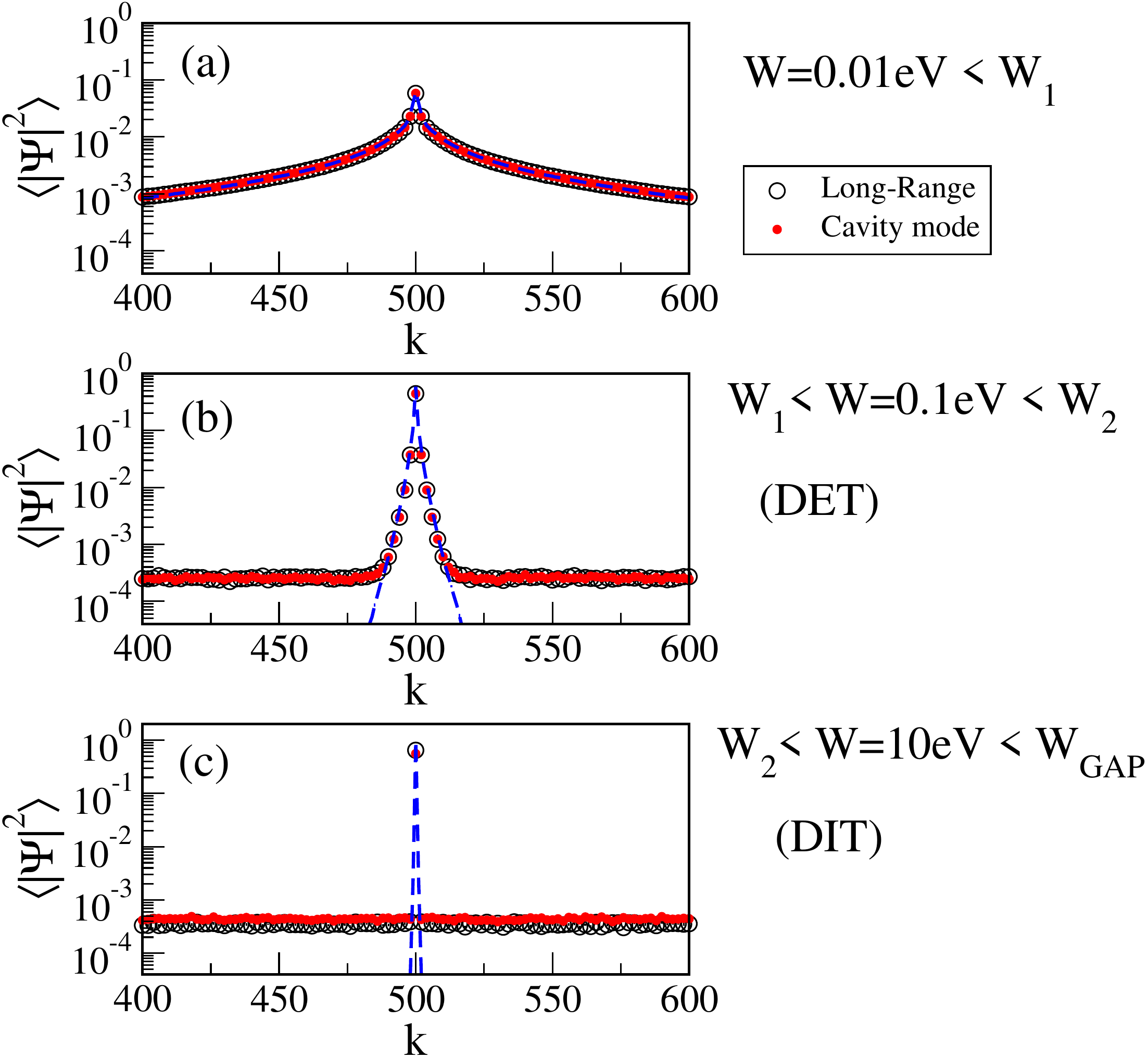}
    \vspace{-2mm}
   \caption{Average shape of the eigenfunctions $\braket{|\Psi|^2}$ in the site basis $k$ comparing the cavity model  (red dots) and the long-range all-to-all coupling $\gamma_{\rm eff}$ (black circles). In each panel different disorder regimes are shown: (a) $ W<W_1$, (b) $W_1 < W<W_2$ and (c) $W_2 < W<W_{\rm GAP}$.     
   Here, $N=10^3$, $\Omega=0.0124$~eV, $\hbar\omega_c=2$~eV, $\mu \approx 36$~D, $g=0.1008$~eV, $g_c=3.188$~eV, $\gamma_{\rm eff}=2g/\sqrt{N}$ and the number of disorder configurations $N_r=10^3$. 
   The blue-dashed lines, shown as a reference, indicate the case $g=0$.}
    \label{Fig:18}
    \end{figure*}

In particular, we note that the zero order matrix $\tilde{H}$ is the same in both cases: the long-range and cavity systems. Moreover, also the mixing $\vec{h}$ between the excited states and the ground state/polaritonic states is the same apart from a factor of $2$. Thus, we can expect that by imposing $\gamma_{\rm eff}=2g/\sqrt{N}$, the excited states of the long-range model and the non-polaritonic states in the molecular chain in the cavity will be very similar until the disorder threshold $W\approx W_{\rm GAP}$, above which disorder will strongly mix the subspaces. 
This hypothesis is verified in Fig.~\ref{Fig:18}, where the average shape of the eigenfunctions $\braket{|\Psi|^2}$ for both the excited states in the long-range case and the non-polaritonic states for the cavity case are compared and shown to be very similar for all the values of disorder considered in the gapped regime. 

\section{Dynamics: wave packet spreading and stationary state}
    
In this Section we analyze the dynamics of a linear chain in presence of long-range hopping and disorder, see  Eq.~(1) in the Main Text. 
We consider a wave packet initially localized on one single site at the center of the chain. We let the initial state evolve in presence of disorder and we compute the variance of the wave packet in time, averaging the probability distribution on the chain sites over the disorder realizations, at each time. 
    
\begin{figure*}[!ht]
    \centering 
    \includegraphics[width=0.65\textwidth]{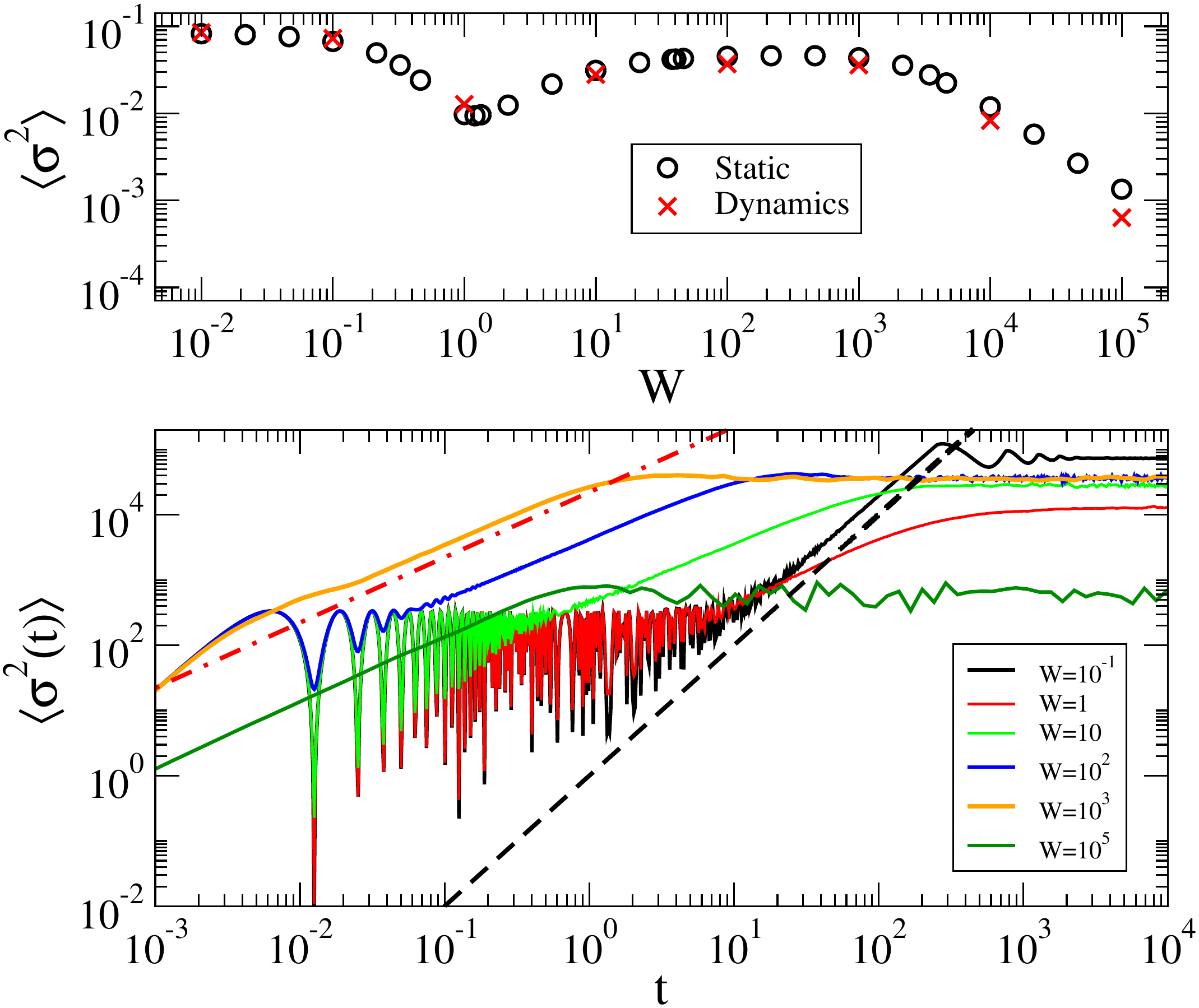}
    \vspace{-3mm}
   \caption{Upper Panel: Average eigenstate variance $\braket{\sigma^2}$ {\it vs.} the disorder strength $W$ of the excited eigenstates (black circles) is compared with the stationary variance (red crosses) obtained by evolving a wave packet initially localized in the middle of the linear chain. The stationary variance has been obtained by averaging over 100 disorder realizations and averaging over time from $500<t<10
  ^4$.  Lower Panel: Variance obtained by evolving a wave packet initially localized in the middle of the liner chain is shown {\it vs.} the time $t$ for different disorder strengths $W$, see legend. At each time the variance has been obtained by averaging over 100 disorder realizations. The ballistic behaviour $\sigma^2 \propto t^2$ is shown as a dashed black line, while the diffusive behaviour $\sigma^2 \propto t$ is shown as a dot-dashed red line.  In both panels a linear chain with long-range hopping has been considered, see Eq.~\eqref{HAcp}. Parameters are: $N=1001,\gamma=\Omega,\Omega=1$ with $W_1 \approx 1.205$, $W_2 \approx 38.126$ and $W_{\rm GAP} \approx 3457.831$. In both panels, the time is measured in units of the hopping time $\hbar/\Omega$.
  }
    \label{Sigma3}
    \end{figure*}

The variance at different times  $\sigma^2 (t)$ has been computed as follows, 
$$
\sigma^2 (t) \equiv \overline{\braket{\psi(t)|x_\alpha^2|\psi(t)}-\braket{\psi(t)|x_\alpha|\psi(t)}^2},
$$
where the over-line stands for the disorder average. The results are shown in Fig.~\ref{Sigma3}(lower panel). Once the variance reaches a stationary value, we computed the time average of the the stationary variance (red crosses) and we plot it for different disorder strengths $W$ in Fig.~\ref{Sigma3}(upper panel), where the average variance of the excited eigenstates is also shown (black circles), see also Fig.~2(b) of the Main Text. Interestingly the two variances, one obtained by analyzing the eigenstates and the other obtained from the dynamics, are very similar. This shows that the average eigenstate variance analyzed in the Main Text can indeed be considered as a figure of merit for transport.  




For some of the red crosses shown in 
Fig.~\ref{Sigma3}(upper panel) we also show the time evolution of the variance $\sigma^2 (t)$ in Fig.~\ref{Sigma3}(lower panel). At small times, one can observe periodic fluctuations independent of the disorder  strength $W$ with frequency $N \gamma/2$ corresponding to the energy gap $\Delta$ between the ground state and the  excited states. At larger times the variance reaches the stationary value in a ballistic-like way $\sigma^2(t) \propto t^2$ for small disorder, see the dashed black line in Fig.~\ref{Sigma3}(lower panel), and in a diffusive-like way $\sigma^2(t) \propto t$  for larger disorder, see the dot-dashed red line in Fig.~\ref{Sigma3}(lower panel). Note that in the disorder-independent transport regime (DIT) $W_2<W<W_{\rm GAP}$, the variance stationary value is independent of the disorder strength $W$.  
For even larger disorder, when $W>W_{\rm GAP}$, the spreading is almost immediately diffusive-like until it saturates.

\begin{figure*}[!ht]
    \centering 
    \includegraphics[width=0.7\textwidth]{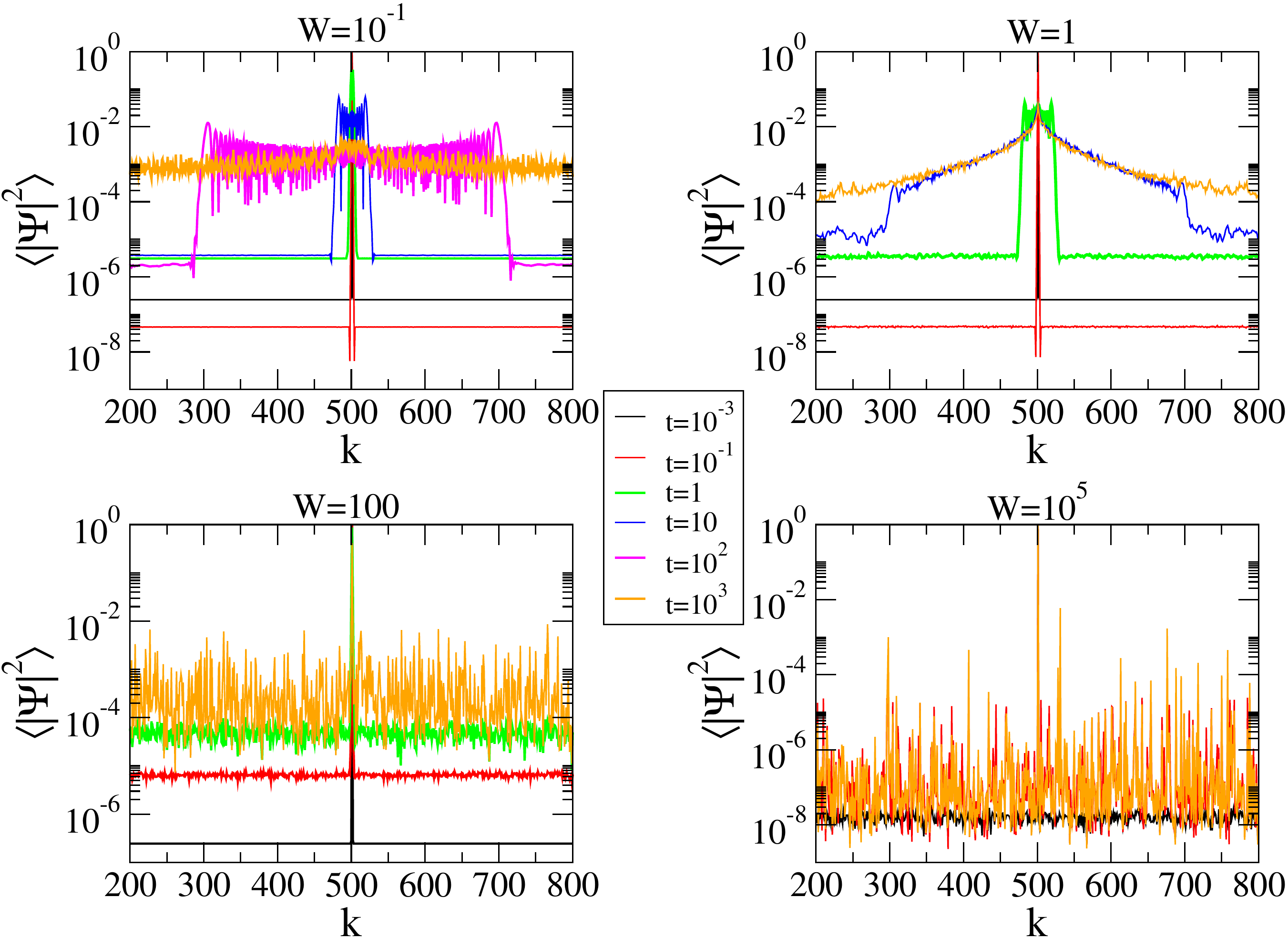}
    \vspace{-4mm}
   \caption{Probability distributions are shown at different times $t$ and different disorder strengths $W$, see legend. The probability distributions have been obtained by evolving an initially localized wave packet at the center of a liner chain. Each probability distribution has been obtained by averaging over 100 disorder realizations. In all panels a linear chain with long-range hopping has been considered, see Eq.~(1) in the Main Text. Parameters are: $N=1001,\gamma=\Omega,\Omega=1$.  In all panels, the time is measured in units of the hopping time $\hbar/\Omega$.}
    \label{Dyn}
    \end{figure*}

While the behaviour of $\sigma^2(t)$ suggests a transition from ballistic to diffusive-like spreading as the disorder strength $W$ increases, a closer look at the probability distribution at different times and for different disorders, Fig.~\ref{Dyn}, shows that both the ballistic and diffusive characterization of the wave packet spreading in presence of long-range hopping are not fully correct. Specifically, for  large $W$,  the increase of the variance is not due to an increase of the width of the initial wave packet but it is mainly due to  the growth of the  flat tails of the probability distribution, see lower panels in Fig.~\ref{Dyn}. 

   \begin{figure*}[!ht]
    \centering 
    \includegraphics[width=0.35\textwidth]{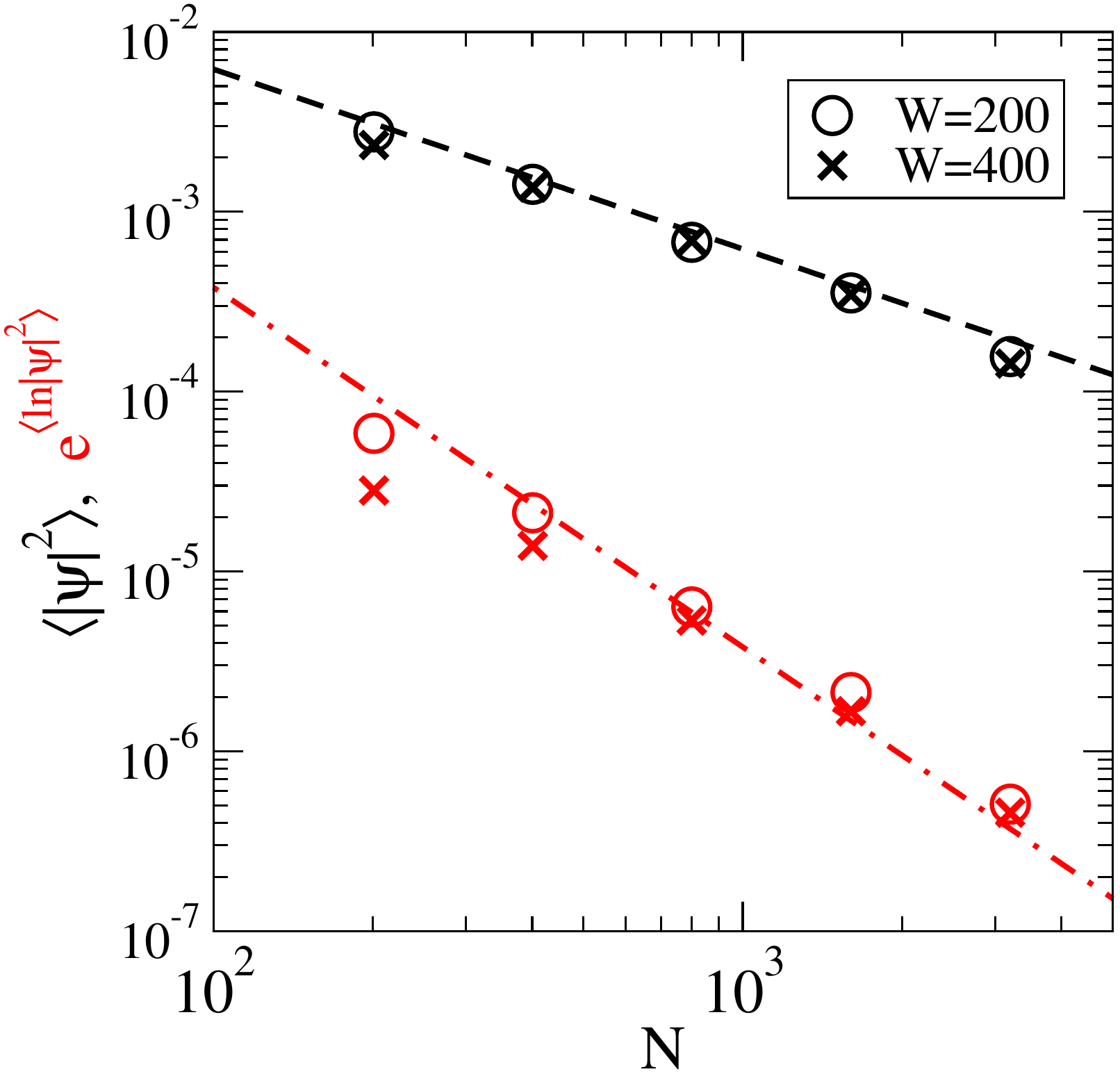}
    \vspace{-2mm}
   \caption{Average  probability in the tails of the stationary distribution is shown for different number of sites $N$ in a linear chain. The average probability in the tails of the stationary distribution has been obtained by evolving a wave packet initially localized at the center of a linear chain. Here a linear chain with long-range hopping has been considered, see Eq.~(1) in the Main Text. Parameters are: $\gamma=\Omega,\Omega=1$ and $W=\{200,400\}$, see legend.  Two different methods to obtain the average probability have been considered:  the average probability $\langle |\psi|
  ^2 \rangle$ (black symbols) and the typical  probability $\exp{\langle \ln{|\psi|^2} \rangle}$ (red symbols) of the tails of the stationary probability distribution (we averaged over all sites but the central one). An additional average over 100 disorder realizations is considered. The black dashed line shows a linear ($1/N$) behaviour while the red dot-dashed line shows a quadratic ($1/N^2$) behaviour.}
    \label{Psi}
    \end{figure*}

Finally, by analyzing the stationary probability distribution obtained from the dynamics we analyzed the tails of the distribution in the disorder-independent transport (DIT) regime. 
Since in this regime the distribution has only one peak at the center of the chain corresponding to the initial state, in order to analyze the statistical properties of the probability distribution in the tails we averaged the probabilities and their logarithm over all the sites but the central one. As one can see from Fig.~\ref{Psi} the average probability in the tails decreases as $\approx 1/N$, while the typical probability 
$e^{\langle \ln{|\psi|^2} \rangle} \approx 1/N^2$, showing that the distribution in the tails is very broad and highly non trivial. Moreover, two different values of the disorder strength $W$ (both in the DIT regime) have been considered,  showing that the tails are independent of disorder in this regime, as discussed above and in the Main Text. 


%